\documentclass[twoside,slac_one]{revtex4}
\usepackage{graphicx}
\usepackage{fancyhdr}
\usepackage{amsmath} % American Mathematics Society standards
\usepackage{bm}% bold math
\usepackage{amsxtra}
\usepackage{amssymb}
\usepackage{amsthm}
\usepackage{latexsym}
\usepackage{lscape}
\RequirePackage{xspace}
\usepackage{epsfig,color,colordvi,amsfonts,multirow,rotating,relsize}
\usepackage{subfigure}

\def\pt        {\mbox{$p_T$}\xspace}
\def\invpb {\ensuremath{\mbox{\,pb}^{-1}}\xspace}
\def\invfb   {\ensuremath{\mbox{\,fb}^{-1}}\xspace}
\def\L{{\ensuremath{\cal L}}\xspace}
\newcommand{\pb}{\ensuremath{\mathrm{pb}}}
\newcommand{\W}{\ensuremath{\mathrm{W}}}
\newcommand{\Wjets}{\ensuremath{\mathrm{W+jets}}}

\newcommand{\WW}{\ensuremath{\W^+\W^-}}
\newcommand{\Z}{\ensuremath{\mathrm{Z}}}

\newcommand{\ZZ}{\ensuremath{\Z\Z}}
\newcommand{\WZ}{\ensuremath{\W\Z}}

\newcommand{\Elp}{\ensuremath{\mathrm{e}^{+}}}
\newcommand{\Elm}{\ensuremath{\mathrm{e}^{-}}}
\newcommand{\Elpm}{\ensuremath{\mathrm{e}^{\pm}}}

\newcommand{\Mp}{\ensuremath{\mu^{+}}}
\newcommand{\Mm}{\ensuremath{\mu^{-}}}

\newcommand{\Mmp}{\ensuremath{\mu^{\mp}}}

\newcommand{\met}{\ensuremath{\Et^{\mathrm{miss}}}}

\newcommand{\Et}{\ensuremath{E_\mathrm{T}}}

\newcommand{\dytt}{\ensuremath{Z/\gamma^*\to\tau\tau}}
\newcommand{\dyll}{\ensuremath{Z/\gamma^*\to\ell\ell}}
\newcommand{\dy}{\ensuremath{Z/\gamma^*}}

\newcommand{\ttbar}{\ensuremath{t\bar{t}}}

\newcounter{myfootertablecounter}

\def\MET {\met}
\begin{document}

%Title of paper
\title{Study of Diboson Production at CMS}

% Repeat the \author .. \affiliation  etc. as needed
%
% \affiliation command applies to all authors since the last
% \affiliation command. The \affiliation command should follow the
% other information

\author{Kalanand Mishra}
\affiliation{Fermi National Accelerator Laboratory, Batavia, IL 60510, USA}

\begin{abstract}
I present an overview of the measurements of the diboson ($WW$, $WZ$, 
$ZZ$, $W\gamma$, and $Z\gamma$) production cross sections 
in proton-proton collisions at $\sqrt{s} = 7$ TeV. 
The measurements are based on 36~\invpb and 1.1 \invfb of data 
collected with the CMS detector at the LHC in 2010 and 2011, respectively. 
The vector bosons $W$ and $Z$ are reconstructed in purely leptonic decays.
The measured cross sections are compared with the Standard Model expectations 
calculated at next-to-leading order in perturbative QCD.  
Limits on anomalous triple gauge boson couplings are derived. 
\end{abstract}

%\maketitle must follow title, authors, abstract
\maketitle

\thispagestyle{fancy}
 
% body of paper here - Use proper section commands
% References should be done using the \cite, \ref, and \label commands
% Put \label in argument of \section for cross-referencing
%\section{\label{}}

%%%%%%%%%%%%%%%%%%%%%%%%%%%%%%%%%%
\section{Introduction}
The gauge boson self-interactions appear as vertices involving three 
or four gauge bosons. 
The study of diboson production in proton-proton collisions is an important 
test of the standard model (SM) because of its sensitivity to the 
self-interaction between gauge bosons via trilinear gauge couplings (TGC).
The values of these couplings are fully fixed in the SM by the gauge 
structure of the $SU(2) \times U(1)$ Lagrangian. 
Any deviation, manifested as an increased cross section,  
would indicate new physics. 
Understanding diboson production is also important for 
Higgs boson searches, because electroweak $WW$ and $ZZ$ production are  
irreducible backgrounds for high mass Higgs. 
%%%%%%%%%%%%%%%%%%%%%%%%%%%%%%%%%%
\section{CMS Detector}
A detailed description of the CMS detector can be found elsewhere~\cite{:2008zzk}. 
The layout comprises a superconducting solenoid providing a uniform magnetic field of 3.8 T.
The bore of the solenoid is instrumented with various particle detection systems.
The inner tracking system is composed of a pixel detector with three barrel layers at radii 
between 4.4 and 10.2 cm and a silicon strip tracker with 10 barrel detection layers extending
outwards to a radius of 1.1\,m. Each system is completed by two end caps, extending 
the acceptance up to $| \eta | < 2.5$.
A lead tungstate crystal electromagnetic calorimeter with fine transverse ($\Delta \eta, \Delta \phi$)
granularity and a brass-scintillator hadronic calorimeter surround the tracking volume and cover the 
region $ | \eta | < 3$.
The steel return yoke outside the solenoid is in turn instrumented with gas detectors
which are used to identify muons in the range $ | \eta | < 2.4$. 
The barrel region is covered by drift tube chambers and the end cap region by cathode strip chambers,
each complemented by resistive plate chambers.
%%%%%%%%%%%%%%%%%%%%%%%%%%%%%%%%%%
\section{ Measurement of the {\boldmath $WW \rightarrow \ell^+ \nu \ell^- \bar{\nu}$} cross section}
%%%%%%%%%%%%
\begin{figure}[!hbtp]
\centering
\subfigure[]{
\centering
\label{subfig:ww_ptmin_0j}
\includegraphics[width=0.4\textwidth]{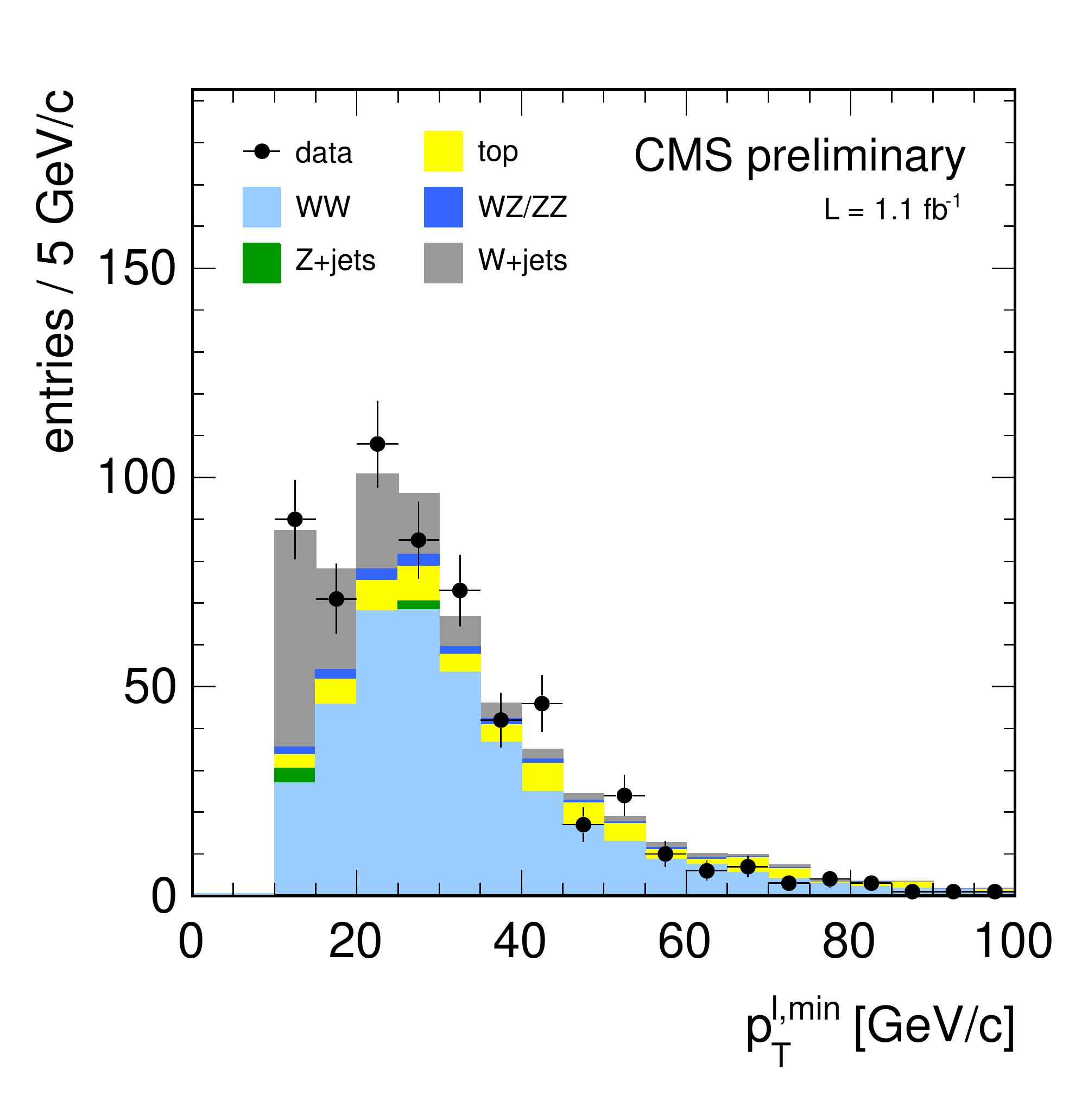}}
\subfigure[]{
\centering
\label{subfig:ww_ptmax_0j}
\includegraphics[width=0.4\textwidth]{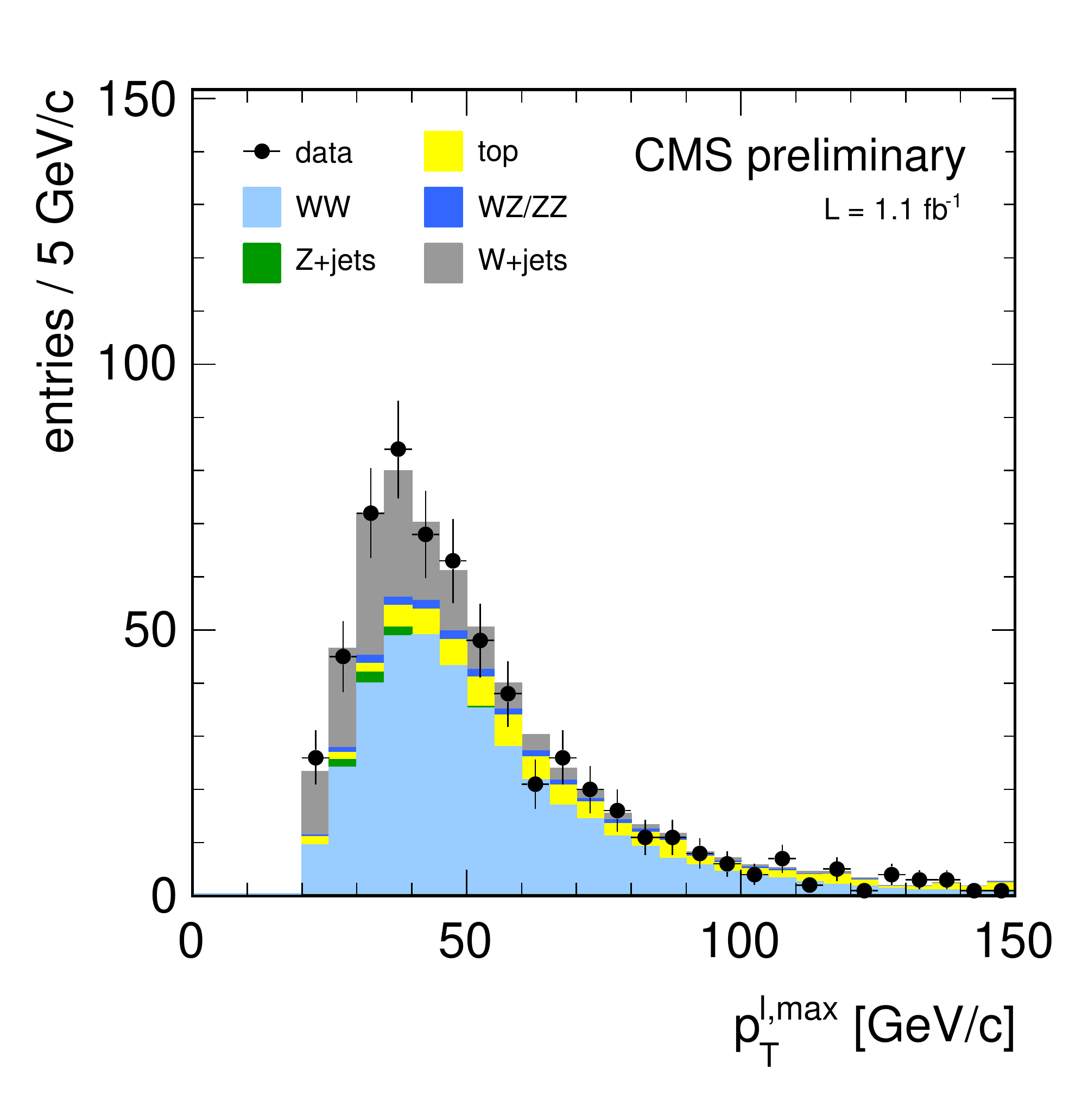}}\\
\subfigure[]{
\centering
\label{subfig:ww_dilmass_0j}
\includegraphics[width=0.4\textwidth]{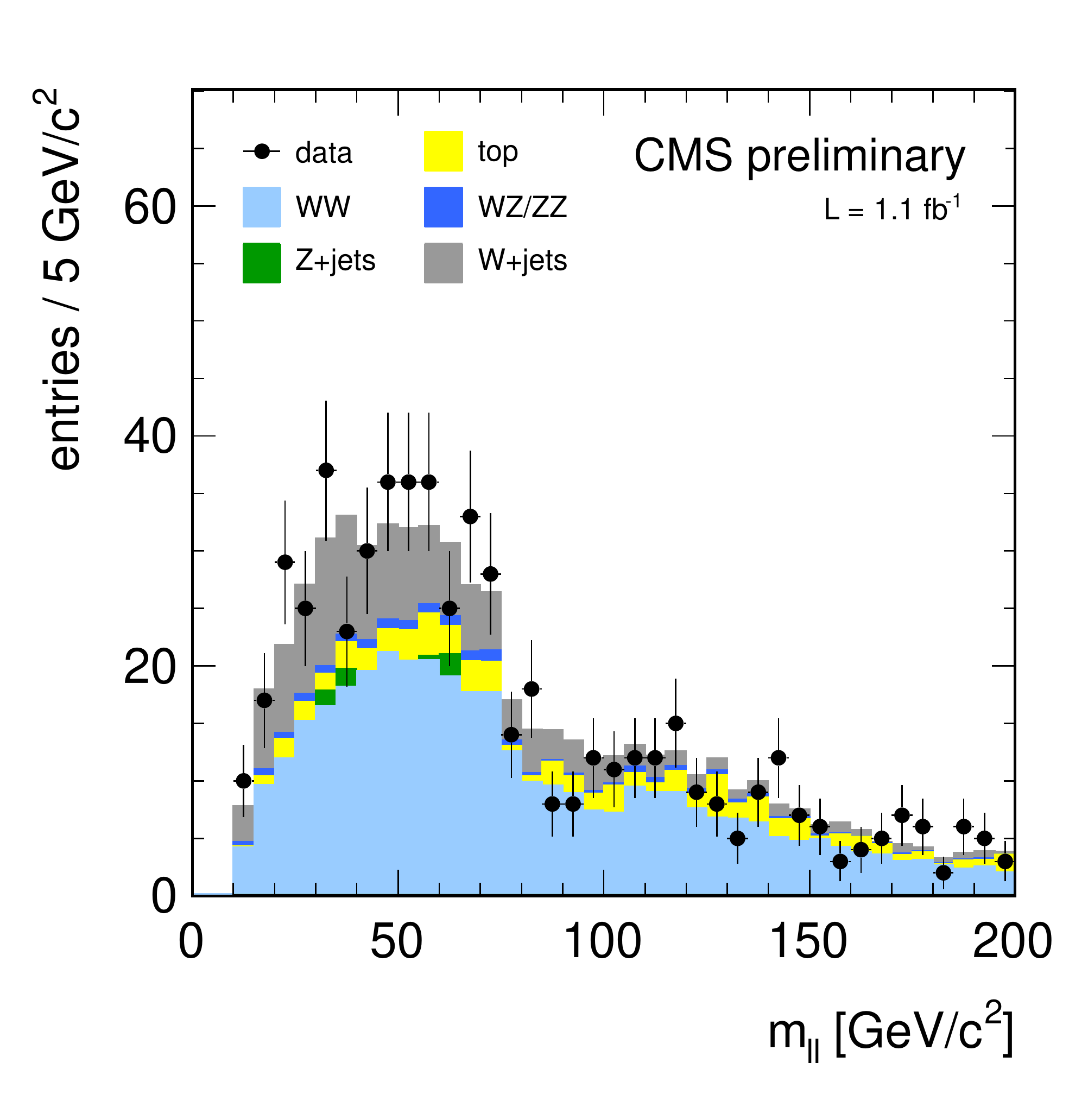}} 
\subfigure[]{
\centering
\label{subfig:ww_pmet_0j}
\includegraphics[width=0.4\textwidth]{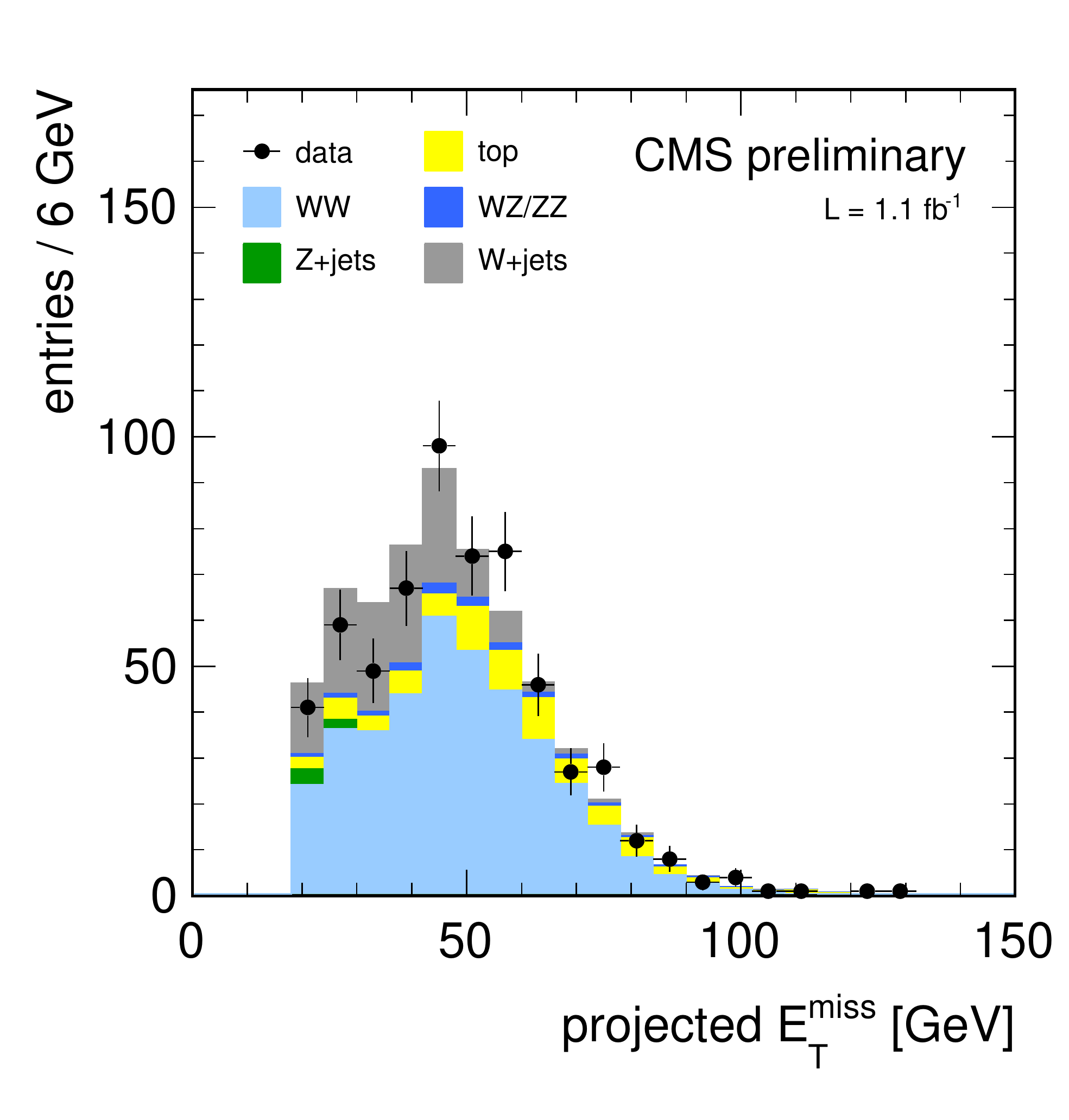}} \\
\caption{$WW$: Distributions of the 
trailing lepton $p_T$ \ref{subfig:ww_ptmin_0j}, leading lepton
$p_T$ \ref{subfig:ww_ptmax_0j}, dilepton invariant
mass \ref{subfig:ww_dilmass_0j} and the
$min(\text{proj}_\text{MET},\text{proj}_\text{trk-MET})$.  Each
component in simulation is scaled to data-driven estimates.}
\label{fig:ww_0j_1}
\end{figure}
%%%%%%%%%%%%%%%%%%%%%%
This measurement is based on data taken in 2011 corresponding
to an integrated luminosity (\L) of 1.1 \invfb.
A similar analysis using 35~\invpb of 2010 data 
is described in Ref.~\cite{Chatrchyan201125}. 
The fully leptonic $\WW$ final state consists of two oppositely charged leptons 
and large missing energy from the two undetectable neutrinos. 
Events are selected using triggers that require the presence of one or two  
high-$\pt$ leptons (electrons or muons). 
Lepton candidates are then reconstructed 
offline and events with two oppositely charged, high-$\pt$, isolated leptons 
($ee$, $\mu\mu$, $e\mu$) are chosen using the following criteria: 
%%%%%%%%%%%%%%%%%%%%%%
\begin{itemize}
\item
Leading lepton $\pt>20$ GeV, second lepton $\pt>10$ GeV.
\item
To reject Drell-Yan events with mismeasured $\met$ associated with poorly
reconstructed leptons, we use the {\it projected $\met$} 
which is the component of $\met$ transverse to the
closest lepton if $\Delta\phi(\ell,\met)<\pi/2$, and
the full $\met$ otherwise.  Events are required to have {\it projected     
$\met$} above 40 GeV in the $\Elp\Elm$ and $\Mp\Mm$ final states,
and above 20 GeV for the $\Elpm\Mmp$ final state.  
These requirements remove more than 99\% of the Drell-Yan background.
\item
To further minimize the Drell-Yan background, events with same flavor 
leptons with a dilepton
invariant mass within $\pm 15$ GeV of the $Z$ mass are rejected. 
Also for this final state,
require $\Delta \phi (\mbox {dilepton, jet}) < 165^{\circ}$ for the 
most energetic jet with
$\pt>15$ GeV to cope with the $Z$+1 jet background. 
\item
Veto events with one or more jets with $\pt>30$ GeV to suppress 
the $W$+jets and top backgrounds.
To further reduce the top quark background, apply a {\it{top veto}} 
based on soft-muon and $b$-jet tagging. 
\item
Background contribution from $ZZ$ and $WZ$ diboson processes is reduced by rejecting
events which have an additional third lepton passing identification and isolation requirements.
\end{itemize}
%%%%%%%%%%%%%%%%%%%%%%

The above steps are described in detail in Ref.~\cite{HWWAnalysis}.
The backgrounds include: \Wjets{} and ${\rm QCD}$ multi-jet events where at 
least one of the jets is misidentified as a lepton, top production (\ttbar{} 
and $tW$), the $\dyll$ process, and other diboson processes  
($WZ$, $ZZ$ and $W\gamma$).
A combination of data-driven methods and detailed Monte Carlo (MC) 
simulation studies are used to estimate background contributions.  
The following backgrounds are estimated from data: $\Wjets$, QCD, 
$\dyll$, top, $\WZ$ and $\ZZ$. 
The remaining background contributions, $W\gamma$ and $\dy\rightarrow\tau\tau$,
are taken from simulation.\\
\begin{table}[!ht]
  \begin{center}
 \caption{Expected number of $\W^+W^-$ signal, and background events from the
   data-driven methods. Uncertainties include both statistical and
   systematic. Signal expectation is taken from simulation assuming NLO cross
   section.}  \label{tab:wwselection_all} 
 \begin{tabular} {l l}
   \hline \hline
   Sample                &   Yield                         \\             \hline
$qq \to \WW $       & 349.7 $\pm$ 30.3     \\
$gg \to \WW $       & 17.2 $\pm$   1.6       \\ \hline
$\Wjets$                & 106.9 $\pm$ 38.9 \\ 
$\ttbar+tW$           &  63.8 $\pm$ 15.9 \\
$\dyll+WZ+ZZ$     & 12.2 $\pm$ 5.3 \\
$\dytt$                  & 1.6 $\pm$ 0.4  \\
$WZ$/$ZZ$ not in $\dyll$ &  8.5 $\pm$ 0.9    \\
$W+\gamma$        & 8.7 $\pm$   1.7  \\ \hline
signal + background  &  569 $\pm$ 52\\ 
Data                       & 626  \\ \hline 
 \end{tabular}
\end{center}
\end{table}
%%%%%%%%%%%%%%%
\begin{table}[!ht]
\begin{center}
\caption{\label{tab:systww} Summary of main systematic uncertainties 
(relative, in \%) in the $W^+W^-$ cross section measurement.}
\vspace{5pt}
\begin{tabular}{l c c c c c c c}
\hline \hline
\multirow{2}{*}{Source} & $qq \to$ & $gg \to$  & non-$\Z$ resonant & top & DY & $\Wjets$ & $V(W/Z)$    \\
                        & $\WW$    & $\WW$       & $\WZ/\ZZ$         &     &         &          & $+\gamma$        \\
\hline
Luminosity                               &  --- & --- &   6 & --- & --- & --- &    6  \\
Trigger efficiencies                     &  1.5 & 1.5 & 1.5 & --- & --- & --- &  1.5  \\
Muon efficiency                          &  1.5 & 1.5 & 1.5 & --- & --- & --- &  1.5  \\
Electron id efficiency                   &  2.5 & 2.5 & 2.5 & --- & --- & --- &  2.5  \\
Momentum scale                           &  1.5 & 1.5 & 1.5 & --- & --- & --- &  1.5  \\
$\met$ resolution                        &  2.0 & 2.0 & 2.0 & --- & --- & --- &  1.0  \\
pile-up                                  &  1.0 & 1.0 & 1.9 & --- & --- & --- &  1.0  \\
Jet counting                             &  5.5 & 5.5 & 5.5 & --- & --- & --- &  5.5  \\  
PDF uncertainties                        &  3.0 & 3.0 & 4.0 & --- & --- & --- &  4.0  \\
\hline
\end{tabular}
\end{center}
\end{table}
%%%%%%%%%%%
%%%%%%%%%%%%%%%%%%%%%%%%%%%%%%%%%%
The total number of expected signal and background events, after applying 
the data-driven corrections, and observed data are
reported in Table~\ref{tab:wwselection_all}.  The distributions of the
key analysis variables are shown in Figure~\ref{fig:ww_0j_1}.
Systematic uncertainties are summarized in Table~\ref{tab:systww}.  
We obtain a total $\WW \to 2\ell 2\nu$ 
efficiency of (6.69 $\pm$  0.51)\%. 
The $\WW$ yield is calculated from the number of events in the signal
region, after subtracting the expected contributions of the various SM 
background processes. From this yield and the $\W \to \ell \nu$
branching fraction~\cite{pdg}, the $\WW$ production
cross section in $pp$ collisions at $\sqrt{s} = $ 7~TeV is found to be
\begin{displaymath}
\sigma_{\WW}  = 55.3 \pm 3.3 \,(\rm{stat}) \pm 6.9 \,(\rm{syst}) \pm 3.3 \,(\rm{lumi})\,\rm{pb}.
\end{displaymath}
This is consistent with the SM expectation of $43.0 \pm
2.0$ pb at NLO~\cite{MCFM} within one standard deviation.
More details on this measurement are given in Ref.~\cite{diboson1p1invfb}.
%%%%%%%%%%%%%%%%%%%%%%%%%%%%%%%%%%%%%%%%%%%%%%%%%%%%%%%%%%%%%%%%%%%%%%%%%%%%%%%%%%%%%%%%
%%%%%%%%%%%%%%%%%%%%%%%%%%%%%%%%%%%%%%%%%%%%%%%%%%%%%%%%%%%%%%%%%%%%%%%%%%%%%%%%%%%%%%%%
%%%%%%%%%%%%%%%%%%%%%%%%%%%%%%%%%%%%%%%%%%%%%%%%%%%%%%%%%%%%%%%%%%%%%%%%%%%%%%%%%%%%%%%%
%%%%%%%%%%%%%%%%%%%%%%%%%%%%%%%%%%%%%%%%%%%%%%%%%%%%%%%%%%%%%%%%%%%%%%%%%%%%%%%%%%%%%%%%
\section{ Measurement of the {\boldmath $WZ\rightarrow\ell\nu\ell^+\ell^-$} cross section}
%%%%%%%%%%%%%%%%%%%%%
\vspace{0.3cm}
\begin{figure}[!htb]
\begin{center}
\includegraphics[width=0.49\textwidth]{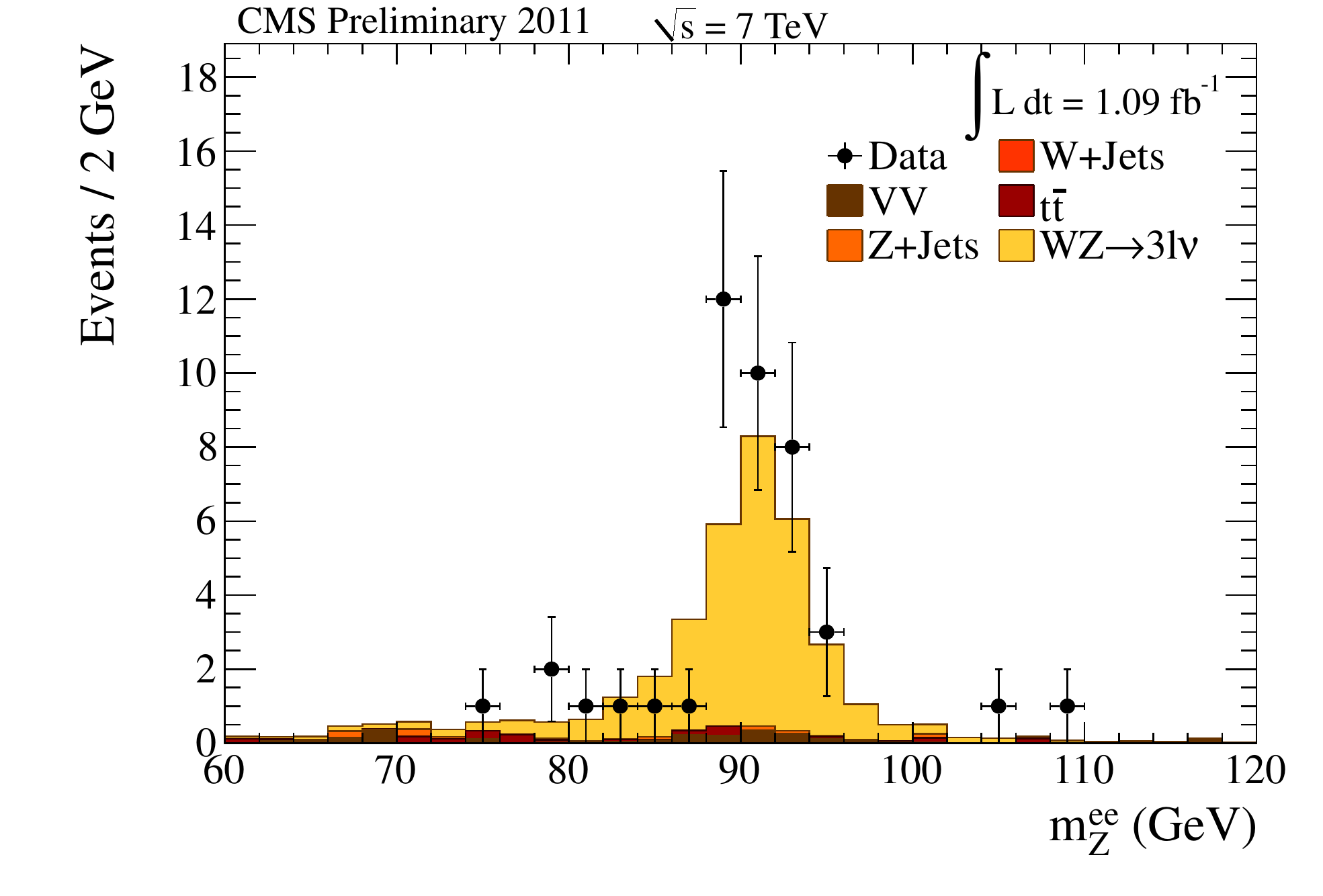}
\includegraphics[width=0.49\textwidth]{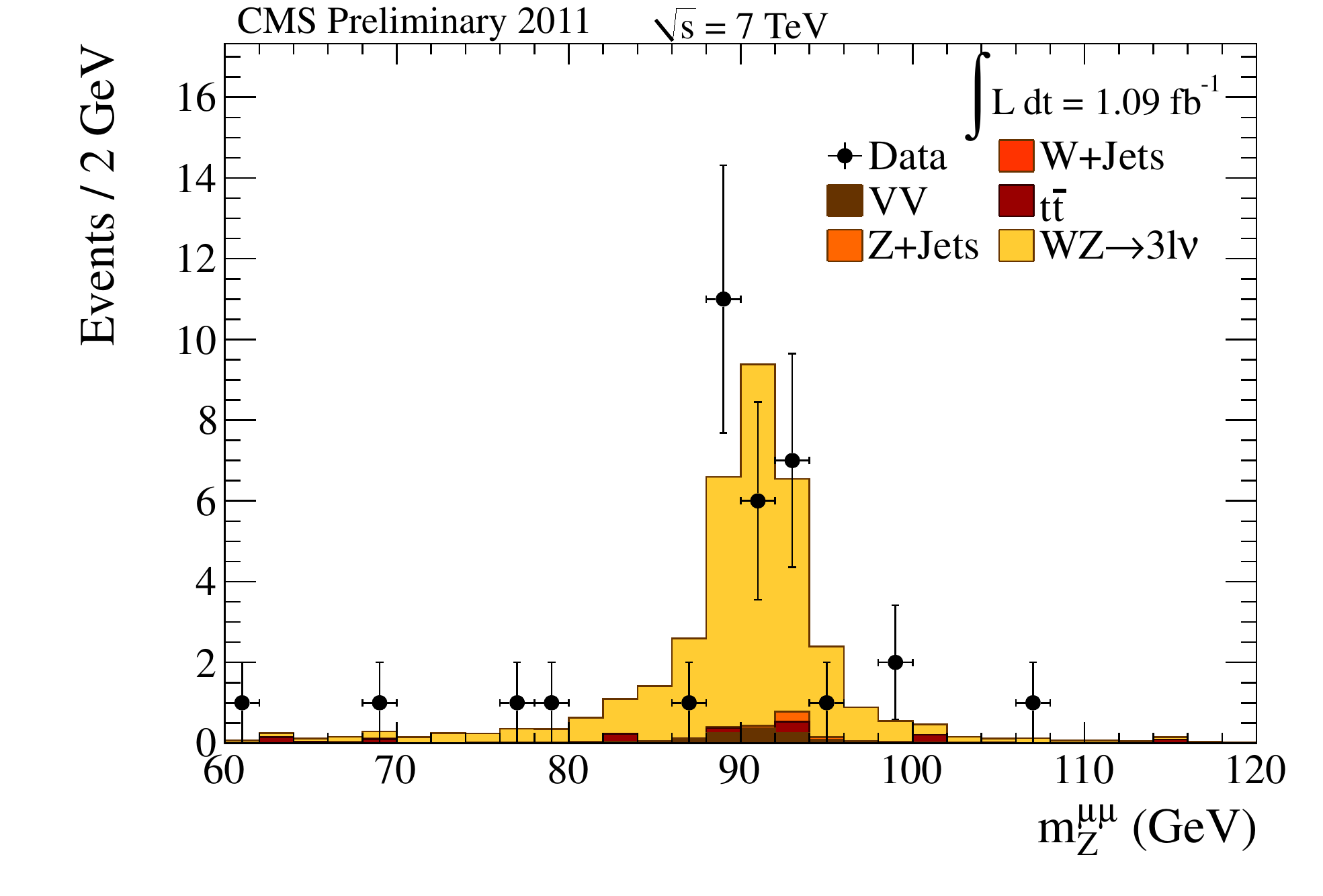}
\caption{$WZ$: Dilepton invariant mass in the $ee$ (left) and 
$\mu \mu$ (right) channels after all selection requirements.}
\label{fig:METcut_Zmass}
\end{center}
\end{figure}
%%%%%%%%%%%%%%%%%%%%%
The $WZ\rightarrow\ell\nu\ell^+\ell^-$ decay is characterized by a pair of same-flavor, opposite-charge
isolated leptons with an invariant mass corresponding to the $Z$ boson, together with a third 
isolated  lepton and large \MET.
The background comes from events with 3 leptons, genuine or fake,
and can be grouped in the following classes:
\begin{itemize}
\item Non-peaking background: di-lepton events without a $Z$ boson, such
  as $t\bar{t}$, QCD or $W$+jets. 
  All but the first of these can be neglected in this analysis.
\item Events with $Z$ + fake lepton, $Z$+jets (including
  $Z$+heavy quarks), or $Z\gamma$ (with photon conversion).
\item Events with a real $Z$ and a third isolated lepton, 
  essentially from $ZZ\to 4\ell$ decays in which one of the four leptons is lost.
  This background is irreducible but is small due to the small $ZZ$ cross section.
\end{itemize}
%%%%%%%%%%%
Candidate events are selected using a double electron or double muon trigger.
The $Z$ boson is  reconstructed from two opposite sign, same flavor leptons 
passing loose identification criteria. 
The leading and second leading lepton are 
required to have $p_T >20 (15)$ GeV and $p_T >10 (15)$ GeV for 
the $Z\rightarrow ee$ ($Z\rightarrow\mu\mu$) case, and their invariant mass
should be in the range 60--120 GeV. 
In case of multiple candidates, the $Z$ candidate with the mass
closest to the nominal $Z$ mass is selected. 
We look for the $W$ boson decay by requiring a third isolated lepton 
with $p_T>20$ GeV, and requiring $\MET$ in the event to be larger than 30 GeV. 
%%%%%%%%%%%%%%%%%%%%%
\begin{table}
  \centering
\caption{Observed and expected number of signal and background yields 
  for the $WZ$ events.}
\label{tab:wzyields}
  \begin{tabular}{l l l l l}
  \hline \hline
  Sample & $3e0\mu$ & $2e1\mu$ & $1e2\mu$ & $0e3\mu$  \\ \hline
   $Z$+Jets & 0.89 & 0.10 & 0.31 & 0.17 \\ 
   $t\bar{t}$ & 0.83 & 0.95 & 0.56 & 0.59\\ 
   $ZZ\rightarrow4\ell$ & 0.40 & 0.95 & 0.40 & 0.97 \\ 
   $V\gamma$ & 0.80 & 0.10 & 0.03 & 0.00 \\ 
   $W$+Jets & 0.00 & 0.00 & 0.00 & 0.00 \\ 
   $WW\rightarrow2\ell2\nu$+Jets & 0.02 & 0.04 & 0.00 & 0.00 \\ \hline
   Total Background & 2.95 & 2.14 & 1.31 & 1.72 \\
   $WZ\rightarrow3\ell\nu$ & 14.47 & 17.49 & 13.95 & 18.56 \\
   All MC & 17.42 & 19.62 & 15.26 & 20.28 \\ 
   Data & 22 & 20 & 13 & 20 \\ \hline 
  \end{tabular}
\end{table}
%%%%%%%%%%%%%%%%%%%%%

The efficiency for leptons to pass the isolation 
and identification requirements is measured using ``tag-and-probe'' 
method from the $Z$ events in data.
The measured efficiency values for muons and electrons are 
$97\%$ and $94\%$, respectively.
In a data sample corresponding to $\L = 1.1~\invfb$, 75 events pass 
these selection criteria.
The data yield and MC expectations for each channel are given 
in Table~\ref{tab:wzyields}. 
The invariant mass of the $Z$ candidates for the selected events
is shown in Fig.~\ref{fig:METcut_Zmass}.
We estimate the $Z$+jets background using the 
data sidebands, and the fake-lepton originated backgrounds by computing the 
jet $\to$ lepton fake rate from $W+$jets events in data.
Similarly, we estimate the $t\bar{t}$ background contamination within 
the signal region using $t\bar{t}$ di-lepton events in data.
We estimate all other background contributions, including  
$ZZ\to 4 \ell$, $Z\gamma$, and $WZ\to l^+l^-l'\nu_{l'} $ where either 
$\ell$ or $\ell'=\tau$ from simulation.

The value of $\mbox{acceptance} \times \mbox{efficiency}$ 
is 19\% for the $eee$ and $\mu\mu e$ final states each, 23\% for the $ee\mu$ 
final state, and 25\% for the $\mu\mu\mu$ final state.
A summary of systematic uncertainties is given in Table~\ref{tab:wzSystematics}.
The cross sections for the four channels are combined, taking into account
the correlations among the systematic uncertainties and 
known branching ratios~\cite{pdg}. 
This results in the cross section measurement
\begin{displaymath}
\sigma( pp \to \mathrm{WZ} + X ) 
= 17.0 \pm 2.4~(\text{stat.}) \pm 1.1~(\text{syst.}) \pm 1.0~(\text{lumi.})~\pb.
\end{displaymath}
The theoretical NLO prediction is $19.79 \pm  0.09$~\cite{MCFM}, 
which is in good agreement with the measured value. 
Cross section measurements in the individual channels are 
consistent with the central value.
More details on this measurement are given in Ref.~\cite{diboson1p1invfb}.
%%%%%%%%%%%%%%%%%%%%%%%%%%%%%%%
\begin{table}[h]
  \begin{center}
    \caption{Summary of systematic uncertainties in the $WZ$ $\to 3 \ell$ cross section measurement.}
    \label{tab:wzSystematics}
    \begin{tabular}{l c l l l l}
      \hline\hline
                                             &                         &$eee$ & $ee\mu$  & $\mu\mu e$  & $\mu\mu\mu$          \\ \hline
      Source                                 & Systematic uncertainty  & \multicolumn{4}{c}{Effect on $\mathcal{F} = A \cdot \epsilon_{MC}$} \\ \hline
      Electron energy scale                  &        2\%              & 1.7\%         & 0.25\%  & 0.9\%  &  n/a    \\
%      Electron energy resolution             &                         &             &        &        &  n/a      \\
      Muon $p_T$ scale                       &        1\%              & n/a         & 0.5\%  & 0.2\%  & 0.9\%      \\ 
%      Muon $p_T$ resolution                  &                         & n/a        &        &        &            \\ 
      MET Resolution                         &                         &   0.5\%     & 0.5\%  &  0.5\% &  0.5\%     \\
      MET Scale                              &                         &   0.3\%     & 0.2\%  &  0.1\% &  0.1\%     \\
      Pileup                                 &                         &   3.1\%     & 0.8\%  &  1.6\% &  1.6\%     \\
      PDF                                    &        1.0\%            & 1.0\%       & 1.0\%  & 1.0\%  & 1.0\%      \\ 
      NLO effect                             &        2.5\%            & 2.5\%       & 2.5\%  & 2.5\%  & 2.5\%      \\ \hline
      Total uncertainty on $\mathcal{F} = A \cdot \epsilon_{MC}$ &     & 4.5\%       & 2.9\%  & 3.3\%  & 3.3\%      \\ \hline 
     Source                                 & Systematic uncertainty  & \multicolumn{4}{c}{Effect on $\rho_{\mbox{eff}}$} \\ \hline
      Electron trigger		             &        1.5\%            & 1.5\% 	     & 1.5\%  & n/a     &  n/a      \\ 
      Electron reconstruction                &        0.9\%            & 2.7\% 	     & 1.8\%  & 0.9\%   &  n/a      \\ 
      Electron ID and isolation              & 2.5\%\ (loose), 3.2\%\ (tight)& 5.9\%       & 5.0\%  & 3.2\%   &  n/a      \\
      Muon trigger                           &        0.54\%           & n/a         & n/a    & 1.08\%  &  1.08\%   \\
      Muon reconstruction                    &        0.74\%           & n/a         & 0.74\% & 1.48\%  &  2.22\%   \\ 
      Muon ID and isolation                  &        0.74\%           & n/a         & 0.74\% & 1.48\%  &  1.94\%   \\ 
     Total uncertainty on $\rho_{eff}$       &                         & 6.7\%       & 5.6\%  & 4.2\%   &  3.6\%    \\ \hline
      Source                                 & Systematic uncertainty  & \multicolumn{4}{c}{Effect on WZ yield} \\ \hline
      Background estimation                  &                         &             &        &         &           \\ \hline
      $ZZ$                                   &        7.5\%             & 0.2\%       & 0.4\%  & 0.3\%   &   0.4\%   \\ 
      $Z\gamma$                              &        13\%             & 0.5\%      & 0.08\% & 0.04\% &   0.08\%  \\ 
      $t\bar{t}$                             &                         & 1.3\%       & 1.3\%  & 0.9\%   & 0.5\%     \\ 
      Jet fake rate                          &                         & 3.3\%       & 4.9\%  & 5.2\%   & 4.2\%    \\ \hline
     Source                                  & Systematic uncertainty  & \multicolumn{4}{c}{Effect on luminosity}   \\ \hline
      Luminosity                             & 6.0\%                   & 6.0\%       & 6.0\%  & 6.0\%   & 6.0\%     \\ \hline
    \end{tabular}
\end{center}
\end{table}
%%%%%%%%%%%%%%%%%%%%%%%%%%%%%%%%%%%%%%%%%%%%%%%%%%%%%%%%%%%%%%%%%%%%%%%%%%%%%%%%%%%%%%%%
%%%%%%%%%%%%%%%%%%%%%%%%%%%%%%%%%%%%%%%%%%%%%%%%%%%%%%%%%%%%%%%%%%%%%%%%%%%%%%%%%%%%%%%%
%%%%%%%%%%%%%%%%%%%%%%%%%%%%%%%%%%%%%%%%%%%%%%%%%%%%%%%%%%%%%%%%%%%%%%%%%%%%%%%%%%%%%%%%
%%%%%%%%%%%%%%%%%%%%%%%%%%%%%%%%%%%%%%%%%%%%%%%%%%%%%%%%%%%%%%%%%%%%%%%%%%%%%%%%%%%%%%%%
\section{ Measurement of the {\boldmath $ZZ\rightarrow\ell^+\ell^-\ell^+\ell^-$} cross section}
%%%%%%%%%%%%%%%%%%%%%%%%%%%%%%%%%%%%%%%%%%%%
\begin{figure}[!htb]
\vspace*{0.3cm}
\begin{center}
\includegraphics[width=0.49\textwidth]{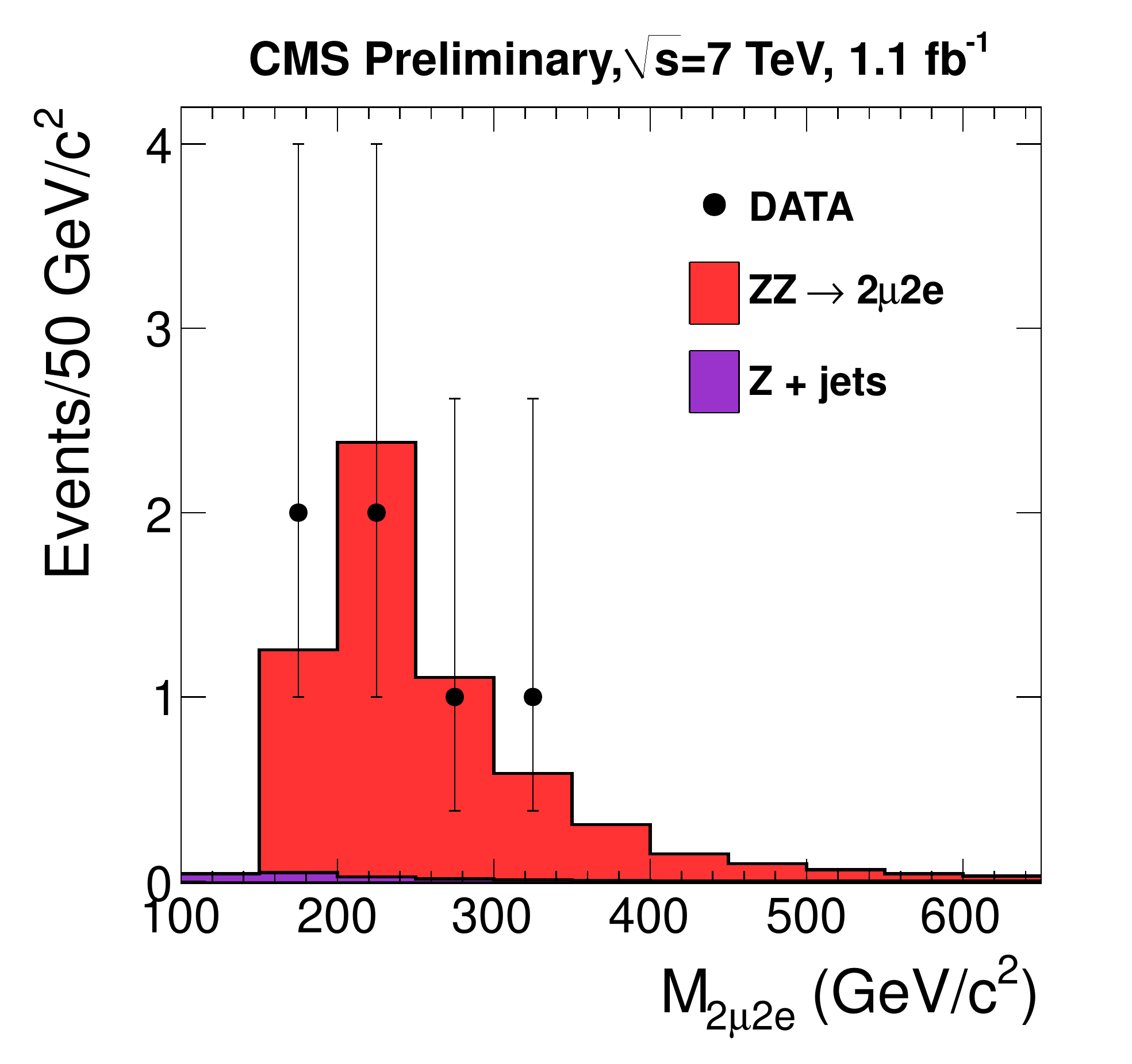} 
\includegraphics[width=0.49\textwidth]{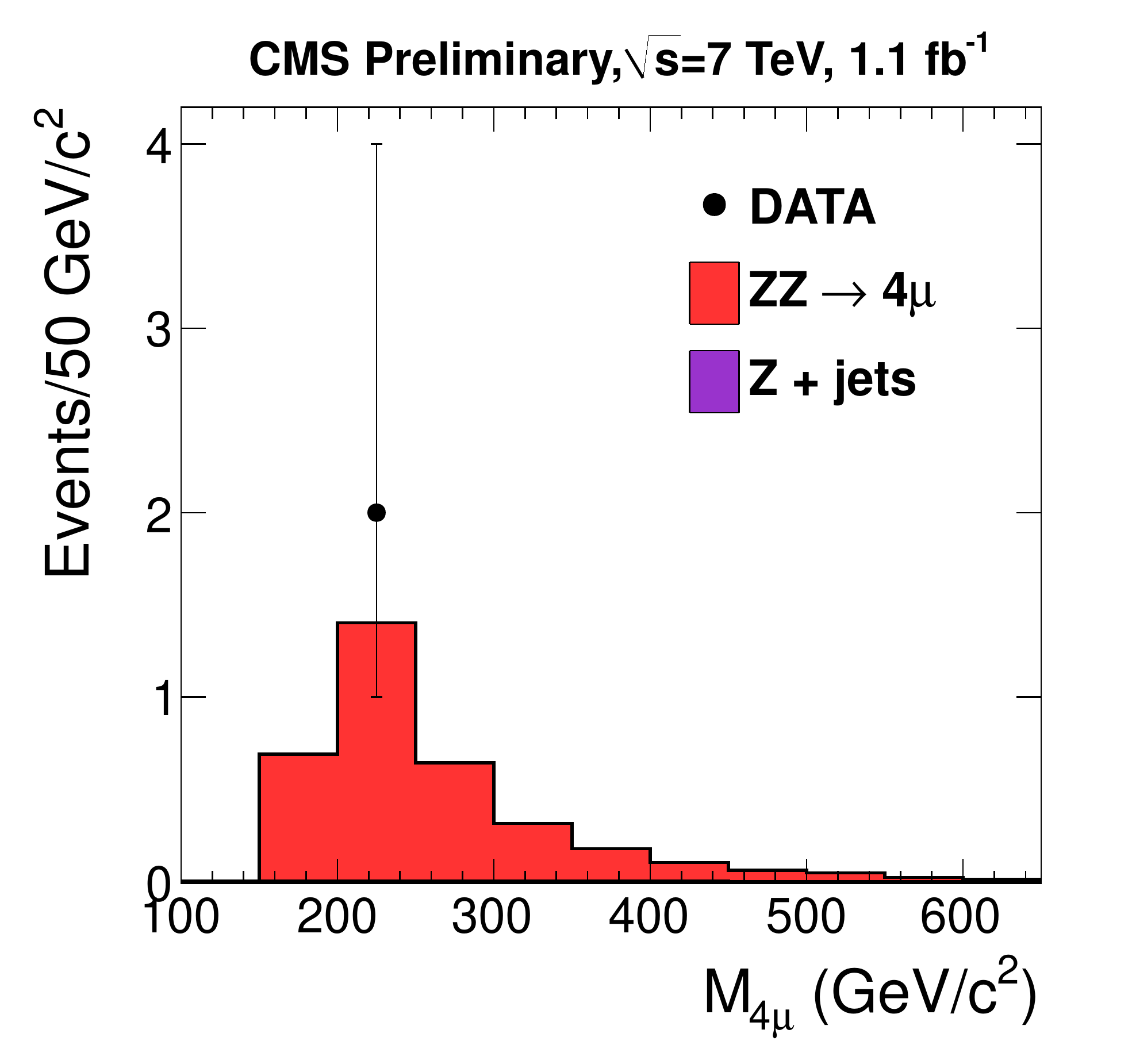} 
\includegraphics[width=0.49\textwidth]{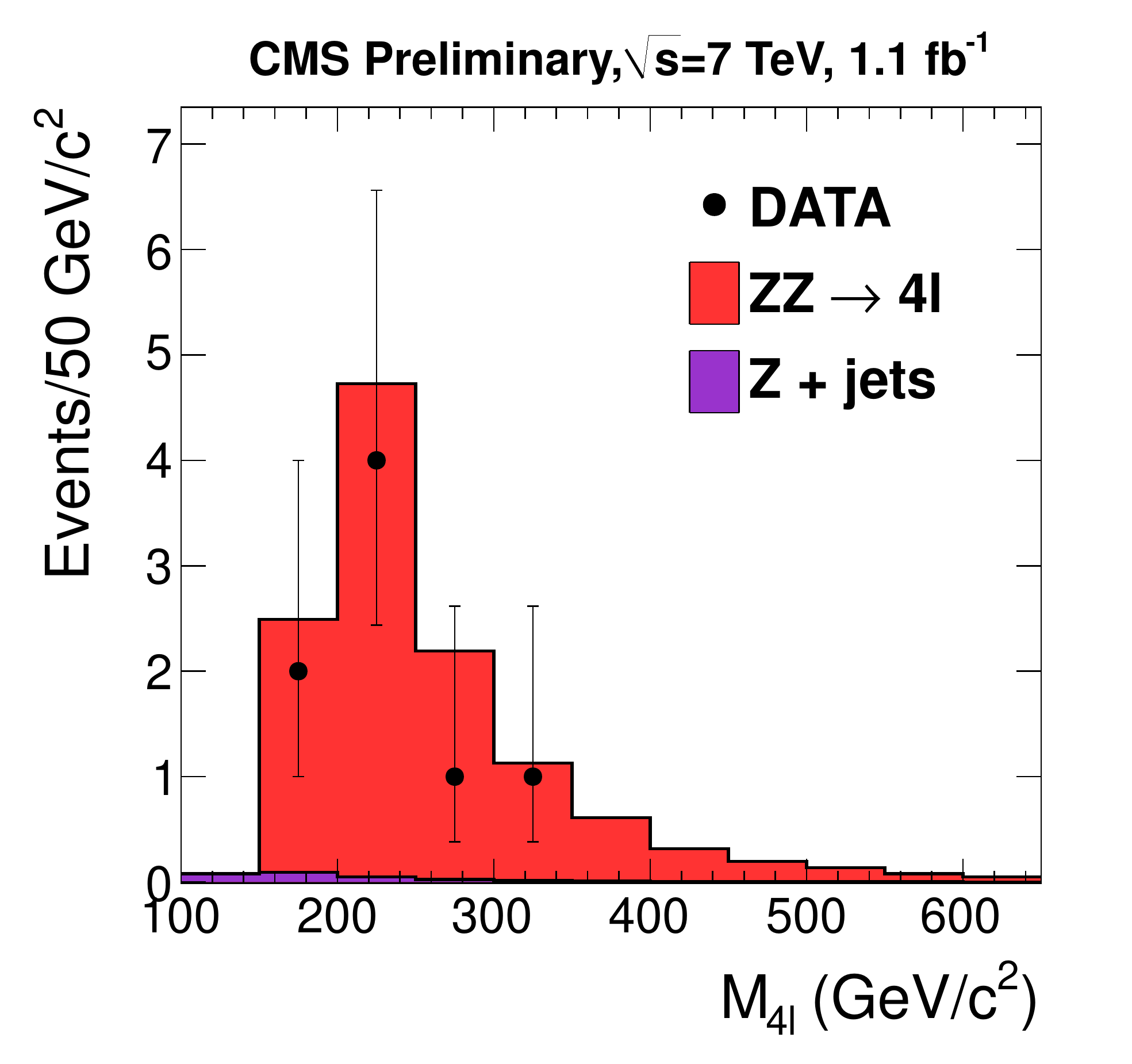} 
\includegraphics[width=0.49\textwidth]{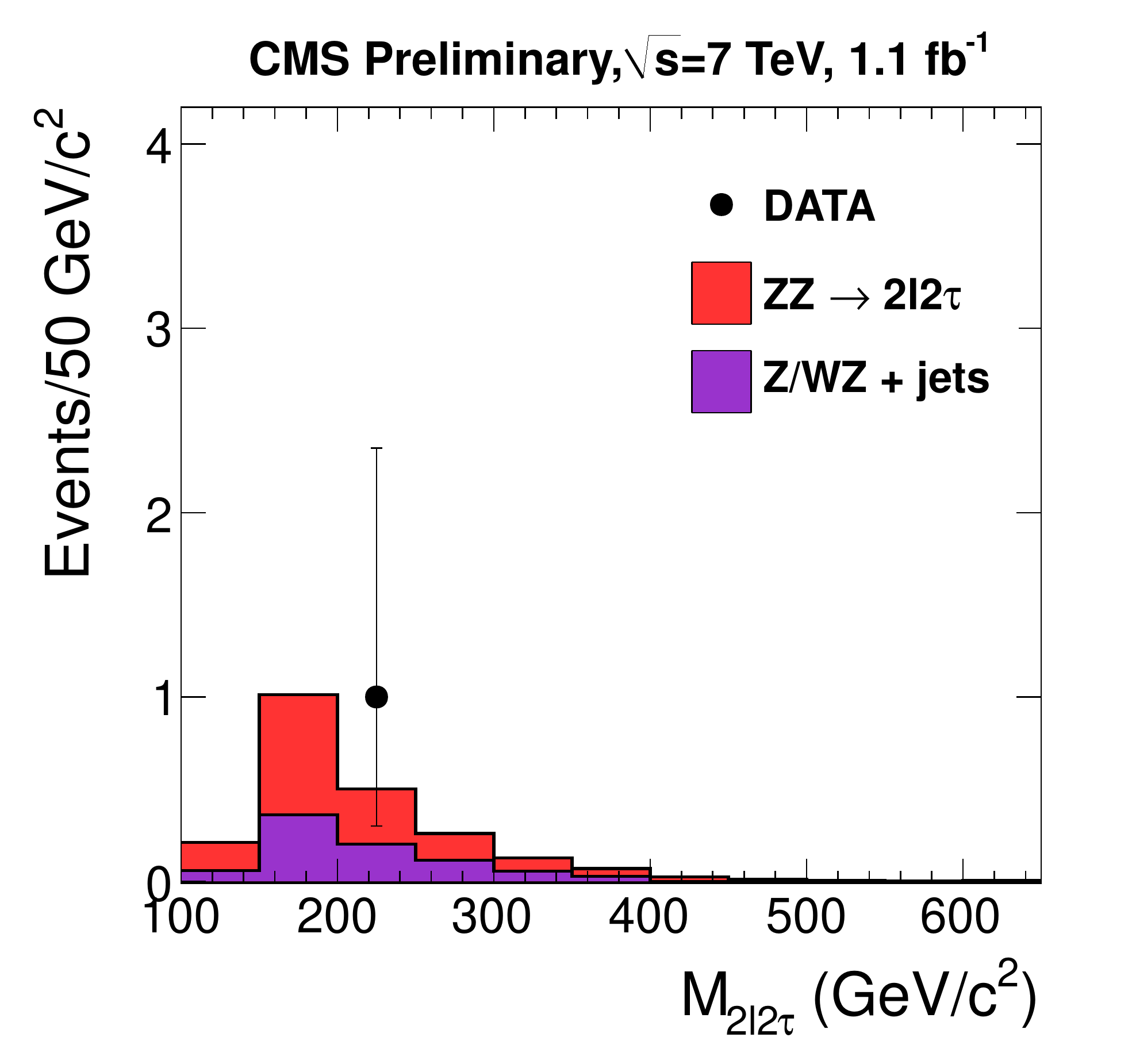} 
\caption{ $ZZ$: Distributions of the four-lepton reconstructed mass for 
the $2e2\mu$ and the $4\mu$ final states (top).
No events were observed in the $4e$ final state.
The bottom left plot represents the sum of the three $4\ell$ channels.
The bottom right plot shows the invariant mass of the $2l2\tau$ final state.
The data samples correspond to $\L = 1.1~\invfb$. }
\label{fig:Mass4lHighmass}
\end{center}
\end{figure}
%%%%%%%%%%%%%%%%
The $ZZ\rightarrow  \ell^{\pm}\ell^{\mp}\ell^{'\pm}\ell^{'\mp}$ process with 
$\ell, \ell' = e$, $\mu$, or $\tau$ is  characterized by two pairs 
of same flavor, opposite charge, high $p_T$, isolated leptons, coming 
from the primary vertex, 
with an invariant mass corresponding to a $Z$ boson.
The process has a clean signature with very little experimental background.
We reconstruct each $Z$ boson in the mass range $60 < m_{\rm Z} < 120$~GeV.
One $Z$ is required to decay into a pair of electrons or muons, and 
the second $Z$ can decay to $\mu\mu, ee$ or $\tau\tau$.
The data sample used for this analysis corresponds to $\L = 1.1~\invfb$, 
and most of the events are selected using a double electron or double muon 
trigger.
For the $4\ell$ final state with $\ell = e,\mu$,  we require the following 
event selection:
\begin{enumerate}
\item {\it First Z:}
         a pair of loosely identified lepton candidates of opposite charge and matching flavor ($e^+ e^-$, $\mu^+\mu^-$) 
         satisfying $m_{1,2} > 60$~GeV, $p_{T,1} > 20$~GeV and $p_{T,2} > 10$~GeV; the pair
         with reconstructed mass closest to the nominal $Z$ boson mass is retained.         
\item {\it Choice of the ``best $4\ell$'':} retain a second lepton pair of opposite charge and matching flavor, 
         among all the remaining $\ell^+ \ell^-$ combinations with $60 < m_{\rm Z} < 120$~GeV and such that 
         the reconstructed four-lepton invariant mass satisfies  $m_{4\ell} > 100$~GeV.
         If more than one combination is found satisfying all the criteria, 
         the one built from leptons of highest $p_T$ is chosen.
\end{enumerate}
%%%%%%%%%%%%%%%%%%%%%%%%
For the $2\ell2\tau$ final state, the first $Z$ boson is required to decay 
to $\mu\mu$ or $ee$ as described above, and the  second $Z$ decays into a 
pair of taus. Each tau candidate can  decay leptonically to a $\mu$ or $e$, or
hadronically. Therefore, there are four possible final states for the 
second $Z$:
$\mu \tau, e \tau, \tau\tau , \mu e$. The selection requirements are: 
%%%%%%%%%%%%%%%%
\begin{itemize}
\item Muon or electron with \pt greater than 10 GeV,
hadronic taus with \pt greater than  20 GeV;
\item The two leptons should be isolated and should have opposite charge.
\item 30 $\le$ Visible Mass ($ll$) $\le$ 80 GeV
\end{itemize}
%%%%%%%%%%%%%%%%

The reducible instrumental background is very small or negligible. 
We estimate any residual background and the associated systematic uncertainty 
using empirical methods based on experimental data.
These are described in more detail in Ref.~\cite{diboson1p1invfb}.
In the $4\ell$ final state, we observe $N_{\rm obs} = 8$ events compared to 
$12.5 \pm 1.1$ events expected from the SM.
The reconstructed four-lepton invariant mass distribution 
is shown in Fig.~\ref{fig:Mass4lHighmass}.
Table~\ref{tab:ZZXS1} shows the number of expected and observed events for 
the individual final states, and also the number of  
background events estimated  using data-driven techniques.
The main sources of systematic uncertainties  are summarized in 
Table~\ref{tab:ZZXS-syst}.
%%%%%%%%%%%%%
\begin{table}[htbp]
\begin{center}
\caption{Number of expected and observed events for the individual $ZZ$ final states.}
\label{tab:ZZXS1}
\begin{tabular}{l l l l} \hline \hline
Final state & $N_{\rm observed}$ & $N^{\rm backg.}_{\rm{estimated}}$ & $N^{ZZ}_{\rm expected}$ \\
\hline
4$\mu$ &  2  & $0.004 \pm 0.004$  & $3.7 \pm 0.4$  \\ 
4$e$   &  0  & $0.14 \pm 0.06$  & $2.5 \pm 0.2$  \\
2e2$\mu$ & 6 & $0.15 \pm 0.06$  &  $6.3 \pm 0.6$ \\  
$2l2\tau$ & 1 & $0.8 \pm 0.1$  & $1.4 \pm 0.1$  \\  \hline
\end{tabular}
\end{center}
\end{table}
%%%%%%%%%%%%%
\begin{table}
\begin{center}
\caption{Summary  of statistical and systematic uncertainties in the $ZZ\to\ell^+\ell^-\ell^+\ell^-$ cross section measurement.} 
\label{tab:ZZXS-syst}
\begin{tabular}{l  l  l  l} \hline \hline
  &  4$\mu$ &  4e   & 2e2$\mu$ \\ \hline 
source  & & { Effects on acceptance A}& \\ \hline
PDF+QCD scale      &  2.2 \% & 2.2 \% & 1.8 \% \\ \hline
source    & \multicolumn{3}{c}{ Effects on efficiency $\epsilon$ } \\ \hline
total uncertainty on $\epsilon$ & 1.7 \% &  3.7 \% & 3.0 \% \\ 
Background (Z+jets)    &  100 \% &  43 \% & 40 \% \\ \hline
Luminosity &  \multicolumn{3}{c}{ 6 \%} \\ \hline 
\end{tabular}
\end{center}
\end{table}
%%%%%%%%%%%%%
The acceptance for the kinematic thresholds and detector coverage is   
in the range 0.56--0.59 for the  4$\mu$, 4$e$ and 2$e$2$\mu$, and
0.18--0.21 for the 2l2$\tau$ final states.
The resulting cross section is
\begin{displaymath}
\sigma (pp \rightarrow {\rm ZZ} + X)  = 3.8 ^{+1.5}_{-1.2} (\rm{stat.}) \pm 0.2 (\rm{sys.}) \pm 0.2 (\rm{lumi.})~\rm{pb}, 
\end{displaymath}
which can be compared to the  theoretical NLO prediction 
${\rm 6.4 \pm 0.6\, pb}$ computed with MCFM~\cite{MCFM}.
More details on this measurement are given in Ref.~\cite{diboson1p1invfb}.
%%%%%%%%%%%%%%%%%%%%%%%%%%%%%%%%%%%%%%%%%%%%%%%%%%%%%%%%%%%%%%%%%%%%%%%%%%%%%%%%%%%%%%%%
%%%%%%%%%%%%%%%%%%%%%%%%%%%%%%%%%%%%%%%%%%%%%%%%%%%%%%%%%%%%%%%%%%%%%%%%%%%%%%%%%%%%%%%%
%%%%%%%%%%%%%%%%%%%%%%%%%%%%%%%%%%%%%%%%%%%%%%%%%%%%%%%%%%%%%%%%%%%%%%%%%%%%%%%%%%%%%%%%
%%%%%%%%%%%%%%%%%%%%%%%%%%%%%%%%%%%%%%%%%%%%%%%%%%%%%%%%%%%%%%%%%%%%%%%%%%%%%%%%%%%%%%%%
\section{ Measurements of the {\boldmath $W\gamma$ and $Z\gamma$} cross sections}
%%%%%%%%%%%%%%%%%%%%%%%%%%%%%%%%%%%%%%%%%%%%
%%%%%%%%%%%%%%%%%%%
\begin{figure}[tb]
  \begin{center}
    \includegraphics[width=0.45\textwidth]{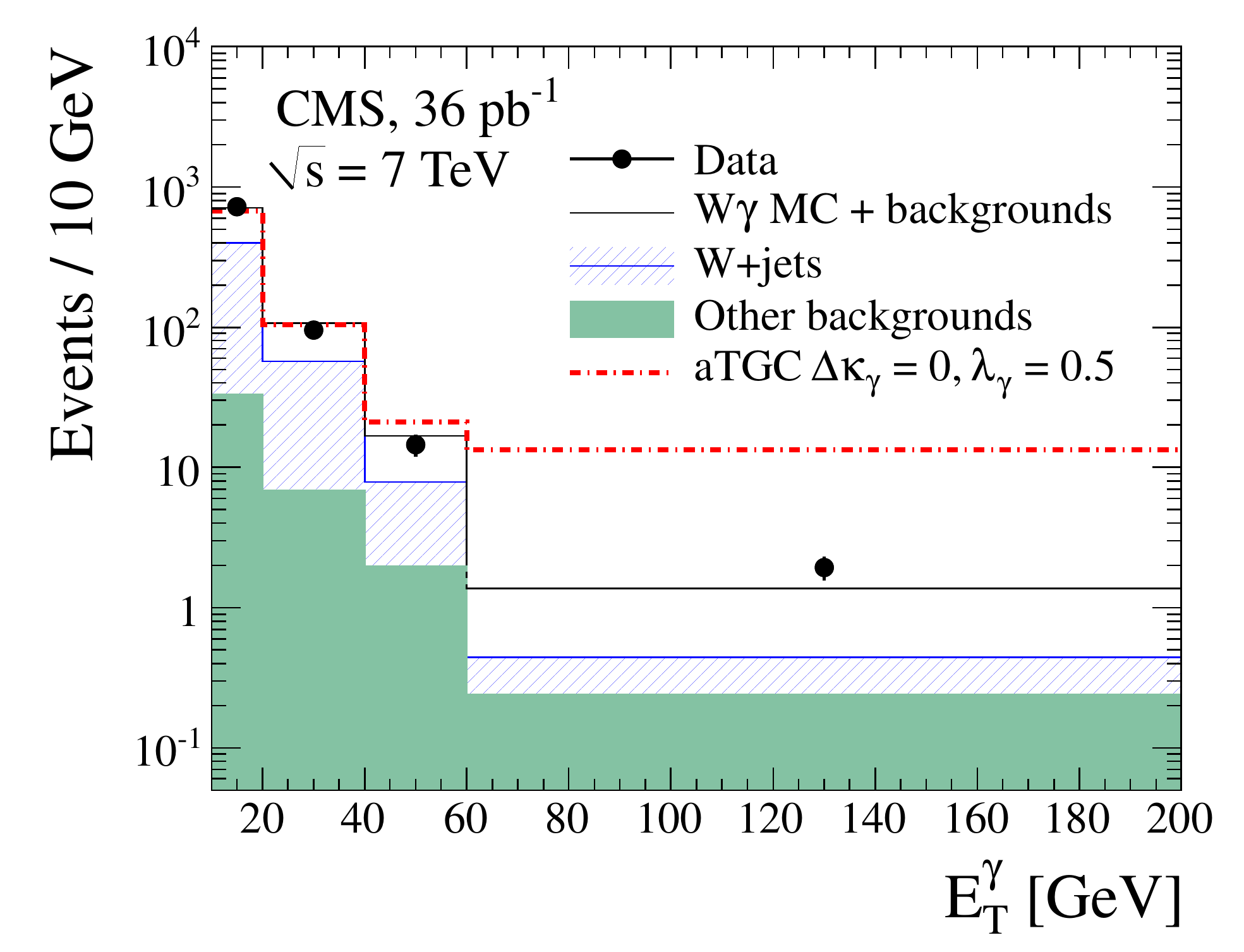}
    \includegraphics[width=0.45\textwidth]{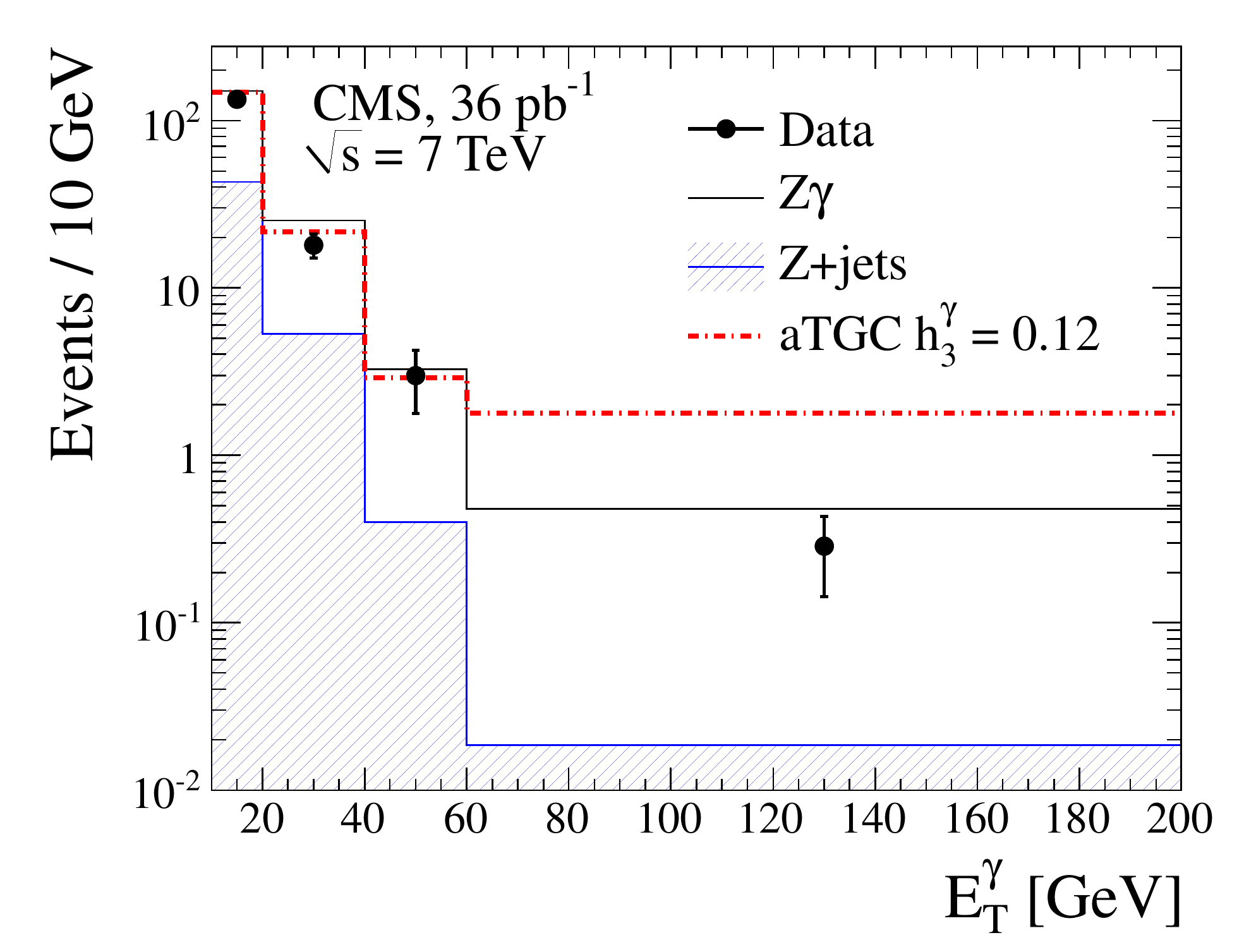}
    \caption{$W\gamma$, $Z\gamma$: 
      Transverse energy distribution of the photon candidate 
      (left: $W\gamma$, right: $Z\gamma$). }
   \label{fig:ETG_wg}
  \end{center}
\end{figure}
%%%%%%%%%%%%%%%%%%%
The $W\gamma \to \ell \nu \gamma $  final state is characterized by a
prompt, energetic, and isolated lepton, significant \MET due
to the presence of the neutrino from the $W$ boson decay, and a prompt
isolated photon.
The $Z\gamma \to \ell \ell \gamma $  final state has two isolated leptons 
and a prompt isolated photon.
Data for this study are selected with a trigger that requires at least 
one energetic electron or muon.
This requirement is about 90\% efficient for the $W\gamma\to\mu\nu \gamma$
signal and 98\% efficient for $W\gamma\to \text{e}\nu \gamma$. The trigger
efficiency is close to 100\% for both $Z\gamma\to \ell\ell\gamma$ final
states. As the $W\gamma$ and $Z\gamma$ cross sections diverge for soft
photons or, in the case of $Z\gamma$ production, for small
values of the dilepton invariant mass, we restrict the cross section measurement
to the phase space defined by photon $E_\text{T} > 10$~GeV and 
$\Delta R(\ell, \gamma)>0.7$.
Furthermore, for $W\gamma$ the \MET in the event must exceed 25 GeV, 
and for the $Z\gamma$ the $m_{\ell\ell}$ must be above 50~GeV.
The data sample used for this analysis corresponds to $\L = 36~\invpb$.

We require a well identified and isolated photon candidate in $|\eta| < 1.44$ or
$1.57 < |\eta| < 2.5$.
The isolated leptons from the $W$ or $Z$ decay are required to have  
$\pt>20$~GeV and $|\eta|<2.5$ (2.4 for muon).
The muon candidate in $W\gamma \to \mu\nu\gamma$
is further restricted to be in $|\eta| < 2.1$.
The main background to $W\gamma$/ $Z\gamma$ production comes from 
$W$/$Z$+jets processes. We estimate this in data by measuring the $E_\text{T}$-dependent
probability for a jet to be identified as photon,
and then folding this probability with the non-isolated photon
candidate $E_\text{T}$ spectrum.
The $E_T$ distribution for photon candidates in events passing the
full selection is given in Fig.~\ref{fig:ETG_wg}.
For $W\gamma$,  we observe  452 events in the $e\nu\gamma$ 
and 520 events in the $\mu\nu\gamma$ final states. The background
from misidentified jets is estimated to be 
$220 \pm 16~\text{(stat.)} \pm 14~\text{(syst.)}$
for $e\nu\gamma$, and $261 \pm 19~\text{(stat.)} \pm 16~\text{(syst.)}$ 
for $\mu\nu\gamma$. Backgrounds from other sources,
such as $Z\gamma$ and diboson, are estimated from simulation and found to be
$7.7 \pm 0.5$ and $16.4 \pm 1.0$ for $W\gamma\to \text{e}\nu\gamma$ and
$W\gamma \to \mu\nu\gamma$, respectively. 
The process $W\gamma \to \tau\nu\gamma$,
with subsequent $\tau\to \ell\nu\nu$ decay, also contributes at the percent
level and is estimated from simulation.
For $Z\gamma$,  we observe 81 events in the $ee\gamma$ and 90 events in the
$\mu\mu\gamma$ final states. 
The $Z$+jets background to these final states is estimated to be
$20.5 \pm 1.7~\text{(stat.)} \pm 1.9~\text{(syst.)}$ and
$27.3 \pm 2.2~\text{(stat.)} \pm 2.3~\text{(syst.)}$, respectively.
Other backgrounds are negligibly small.
All systematic uncertainties are summarized in Table~\ref{tab:systematics}.
%%%%%%%%%%%%%%%%%%%
\begin{figure}[p!]
  \begin{center}
    \includegraphics[width=0.45\textwidth]{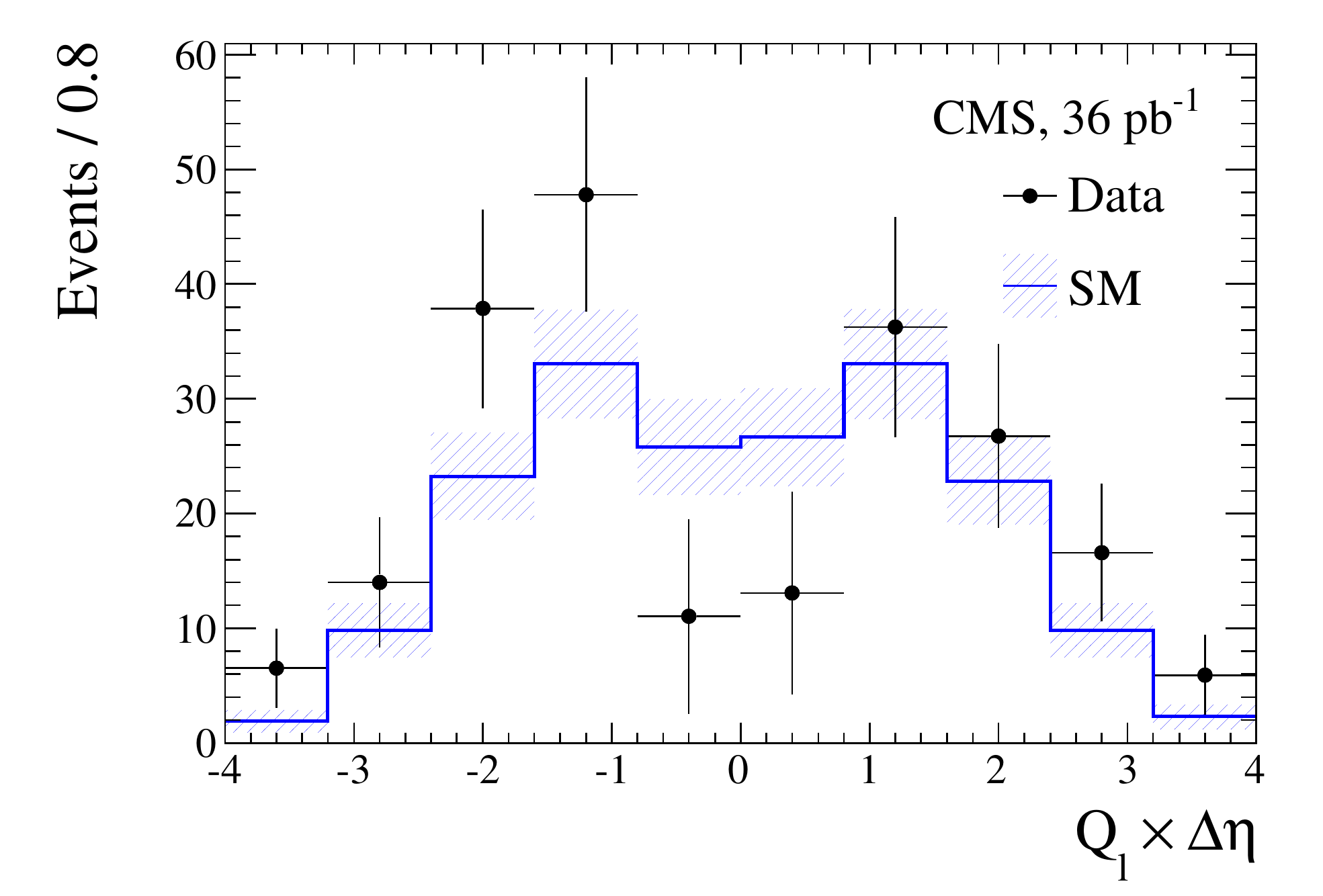}
    \includegraphics[width=0.45\textwidth]{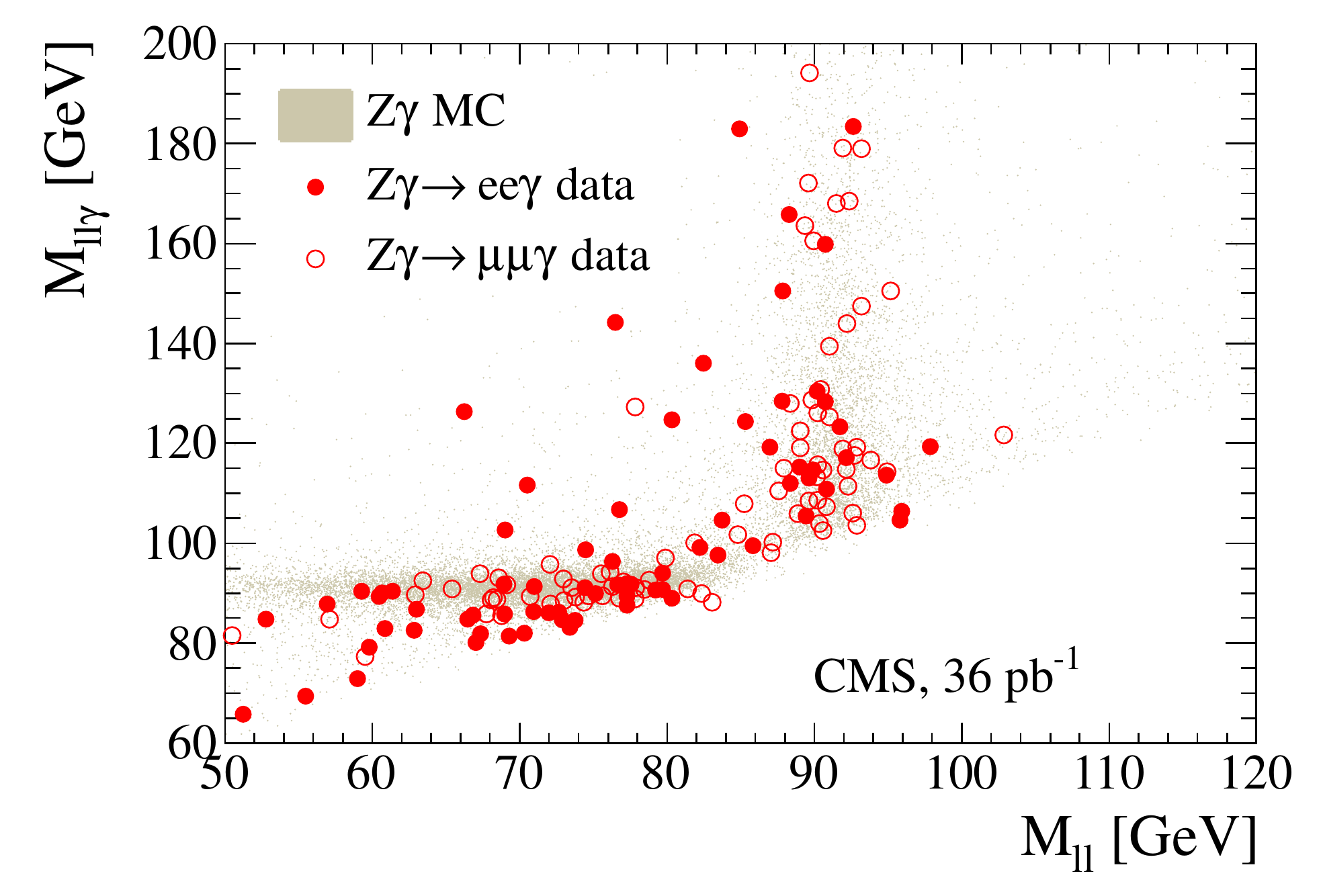}
    \caption{$W\gamma$, $Z\gamma$: (left) Charge-signed rapidity
    difference in the $W\gamma$ production (electron and muon channels combined).
    (right) Distribution of the $\ell\ell\gamma$ invariant vs. 
    $\ell\ell$ invariant mass for the $Z\gamma$ candidates. The data
    accumulation at $M_{\ell\ell\gamma} \simeq M_Z$ corresponds to FSR events, 
    while the data at $M_{\ell\ell} \simeq M_Z$ correspond to ISR events.
    }
    \label{fig:raz}
  \end{center}
\end{figure}
%%%%%%%%%%%%%%%%%%%

Leading order $W\gamma$ production can be described by three
processes: initial state radiation (ISR), where a photon is radiated by
one of the incoming quarks; final state radiation (FSR), where a photon 
is radiated
from the charged lepton from the $W$ boson decay; and finally through the
$WW\gamma$ vertex, where a photon couples directly to the $W$ boson.
The three tree-level $W\gamma$ production processes interfere
with each other, resulting in a radiation-amplitude zero (RAZ) in
the angular distribution of the photon.
In the SM, the location of the dip minimum is located at
$Q_\ell \times \Delta \eta = 0$.
Anomalous $W\gamma$ production can result in a flat distribution.
In Fig.~\ref{fig:raz} we plot the charge-signed rapidity difference in
background-subtracted data. 
There is a good agreement between the data and MC prediction. 
In the SM, leading order $Z\gamma$ production is described via ISR and FSR
processes only, because the $ZZ\gamma$ and $Z\gamma\gamma$
TGCs are not allowed at the tree level.
The distribution of the $\ell \ell \gamma$ mass
as a function of the dilepton mass is shown in Fig.~\ref{fig:raz}.
%%%%%%%%%%%%%%%%%%%
\begin{table}[h]
  \begin{center}
   \caption{Summary of systematic uncertainties in $W\gamma$ and $Z\gamma$ cross section measurements.}
    \label{tab:systematics}
    \begin{tabular}{l c c c c}
      \hline\hline
                                      & $W\gamma\to\ e\nu\gamma$ &  $W\gamma\to\mu\nu\gamma$ & $Z\gamma\to ee\gamma$ & $Z\gamma\to\mu\mu\gamma$   \\ \hline
      Source                                         & \multicolumn{4}{c}{Effect on $A \cdot \epsilon_{\mathrm{MC}}$}   \\ \hline
      Lepton energy scale                            & 2.3\%            & 1.0\%           & 2.8\%      & 1.5\% \\
      Lepton energy resolution                       & 0.3\%            & 0.2\%           & 0.5\%      & 0.4\% \\
      Photon energy scale                            & 4.5\%            & 4.2 \%          & 3.7\%      & 3.0\% \\
      Photon energy resolution                       & 0.4\%            & 0.7\%           & 1.7\%      & 1.4\% \\
      Pile-up                                        & 2.7\%            & 2.3\%           & 2.3\%      & 1.8\% \\
      PDFs                                           & 2.0\%            & 2.0\%           & 2.0\%      & 2.0\% \\ \hline
      Total uncertainty on $A \cdot \epsilon_\mathrm{MC}$   & 6.1\%            & 5.2\%           & 5.8\%      & 4.3\% \\ \hline
                                                     & \multicolumn{4}{c}{Effect on $\epsilon_{\mathrm{data}}/\epsilon_{\mathrm{MC}}$} \\ \hline
      Trigger				             & 0.1\% 	        & 0.5\%           & $<0.1\%$   & $<0.1\%$ \\
      Lepton identification and isolation            & 0.8\% 	 	& 0.3\%           & 1.1\%      & 1.0\% \\
      \MET selection                                 & 0.7\%            & 1.0\%           & N/A        & N/A   \\
      Photon identification and isolation            & 1.2\%            & 1.5\%           & 1.0\%      & 1.0\% \\
      Total uncertainty on $\epsilon_\mathrm{data}/\epsilon_\mathrm{MC}$
                                                     & 1.6\%            & 1.9\%           & 1.6\%      & 1.5\% \\ 
      Background                                     & 6.3\%            & 6.4\%           & 9.3\%      & 11.4\% \\  \hline
      Luminosity  		                     & \multicolumn{4}{c}{ 4\%} \\ \hline
    \end{tabular}
  \end{center}
\end{table}
%%%%%%%%%%%%%%%%%%%

We find the cross section for $W\gamma$ production to be
$\sigma(\mathrm{pp}\to W\gamma+X) \times \mathcal{B}(\text{W}\to \text{e}\nu) =
        57.1  \pm 6.9 \text{~(stat.)}
         \pm 5.1 \text{~(syst.)}
        \pm 2.3 \text{~(lumi.)}$~pb and
$\sigma(\mathrm{pp}\to W\gamma+X) \times \mathcal{B}(\text{W}\to\mu\nu) =
        55.4 \pm 7.2\text{~(stat.)} \pm 5.0 \text{~(syst.)}
        \pm 2.2 \text{~(lumi.)}$~pb.
The combination of the two results yields
$\sigma(\mathrm{pp}\to W\gamma+X) \times \mathcal{B}(\text{W}\to \ell\nu) =
56.3 \pm 5.0~\text{(stat.)} \pm 5.0~\text{(syst.)} \pm 2.3~\text{(lumi.)}$~pb.
This result agrees well with the NLO prediction~\cite{baurWg} of $49.4\pm3.8$~pb.
The Z$\gamma$ cross section is measured to be
$\sigma(\mathrm{pp}\to\text{Z}\gamma+X) \times \mathcal{B}(\text{Z} \to \text{ee}) =
9.5 \pm 1.4~(\text{stat.}) \pm 0.7~(\text{syst.}) \pm 0.4~(\text{lumi.})$~pb
for the ee$\gamma$ final state, and
$\sigma(\mathrm{pp}\to\text{Z}\gamma+X) \times \mathcal{B}(\text{Z}\to \mu\mu) =
9.2 \pm 1.4~(\text{stat.}) \pm 0.6~(\text{syst.}) \pm 0.4~(\text{lumi.})$~pb
for the $\mu\mu\gamma$ final state. The combination of the two results yields
$\sigma(\mathrm{pp}\to\text{Z}\gamma+X) \times \mathcal{B}(\text{Z}\to \ell\ell) =
9.4 \pm 1.0~(\text{stat.}) \pm 0.6~(\text{syst.}) \pm 0.4~(\text{lumi.})$~pb.
The theoretical NLO prediction~\cite{baur} is $9.6\pm0.4$~pb, which is in
agreement with the measured value.
More details on these measurements are given in Ref.~\cite{VGammaCMS}.
%%%%%%%%%%%%%%%%%%%%%%%%%%%%%%%%%%%%%%%%%%%%%%%%%%%%%%%%%%%%%%%%%%%%%%%%%%%%%%%%%%%%%%%%
%%%%%%%%%%%%%%%%%%%%%%%%%%%%%%%%%%%%%%%%%%%%%%%%%%%%%%%%%%%%%%%%%%%%%%%%%%%%%%%%%%%%%%%%
%%%%%%%%%%%%%%%%%%%%%%%%%%%%%%%%%%%%%%%%%%%%%%%%%%%%%%%%%%%%%%%%%%%%%%%%%%%%%%%%%%%%%%%%
%%%%%%%%%%%%%%%%%%%%%%%%%%%%%%%%%%%%%%%%%%%%%%%%%%%%%%%%%%%%%%%%%%%%%%%%%%%%%%%%%%%%%%%%
\section{ Limits on anomalous triple gauge couplings from {\boldmath 36~\invpb} data}
%%%%%%%%%%%%%%%%%%%%%%%%%%%%%%%%%%%%%%%%%%%%
%%%%%%%%%%%%%%%%%%%%
\begin{figure}[tb]
\begin{center}
\includegraphics[width=0.49\textwidth]{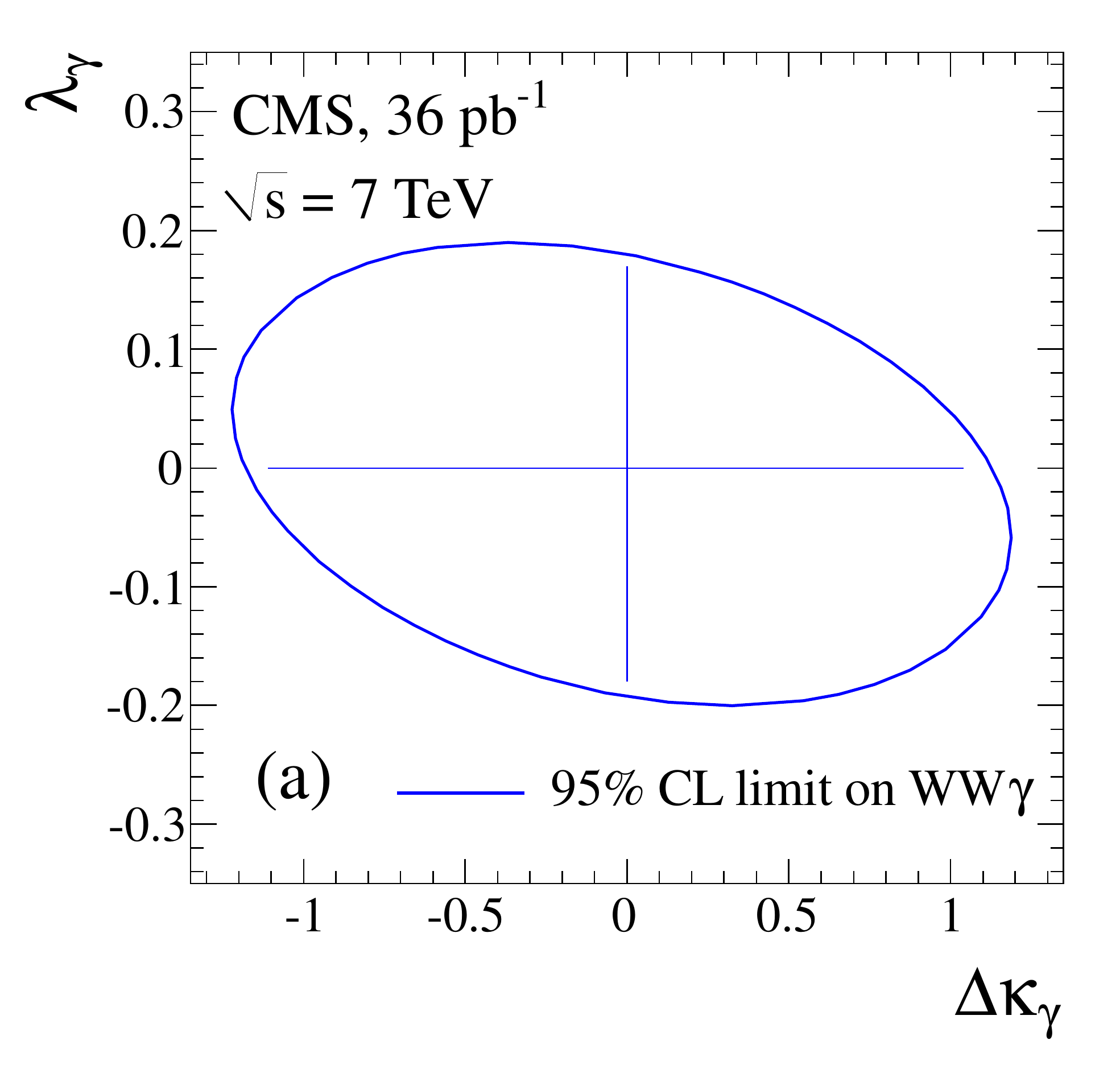}
\includegraphics[width=0.49\textwidth]{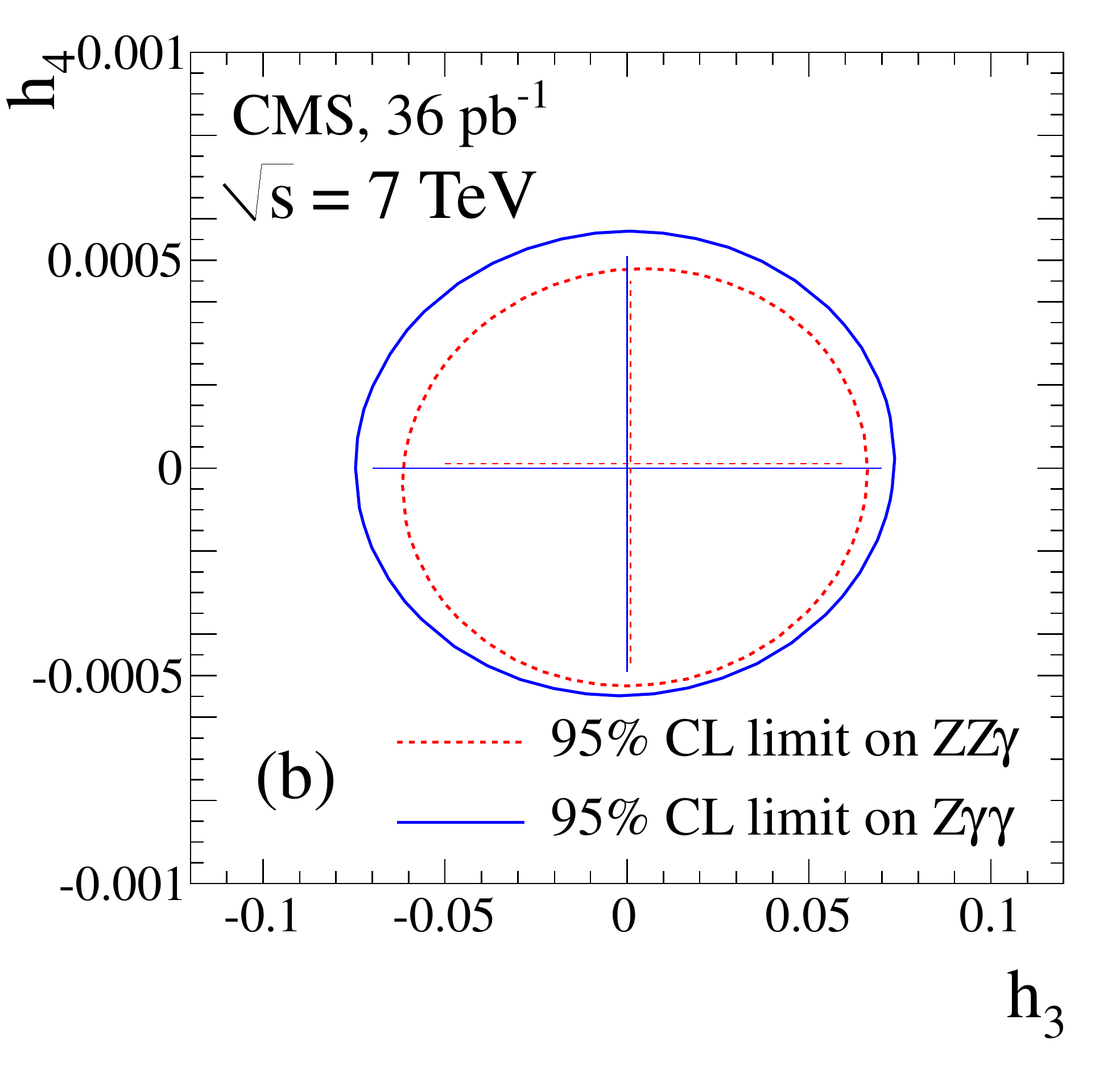}
\caption{Two-dimensional 95\% CL limit contours (a) for the $WW\gamma$ vertex couplings
$\lambda_\gamma$ and $\Delta\kappa_\gamma$ (blue line), and (b) for the
$ZZ\gamma$ (red dashed line) and $Z\gamma\gamma$ (blue solid line) vertex
couplings $h_3$ and $h_4$ assuming no energy dependence on the couplings.
One-dimensional 95\% CL limits on individual couplings are given as
solid lines.}
   \label{fig:atgc_contour}
  \end{center}
\end{figure}
%%%%%%%%%%%%%%%%%%%%
%%%%%%%%%%%%%%%%%%%%%%
\begin{figure}[tp]
  \centerline{
    \subfigure[]{\includegraphics[width=0.49\textwidth]{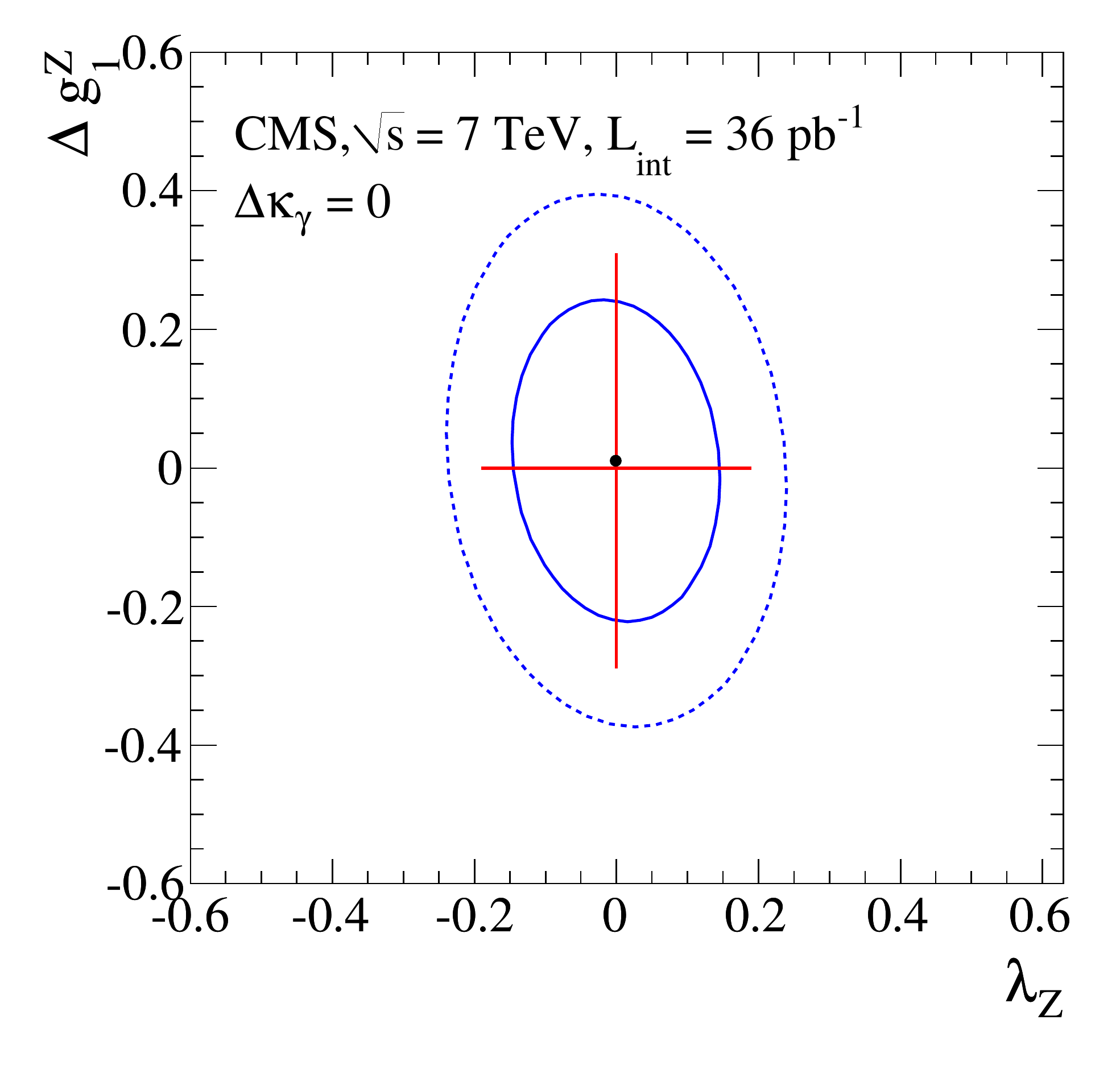}}
    \subfigure[]{\includegraphics[width=0.49\textwidth]{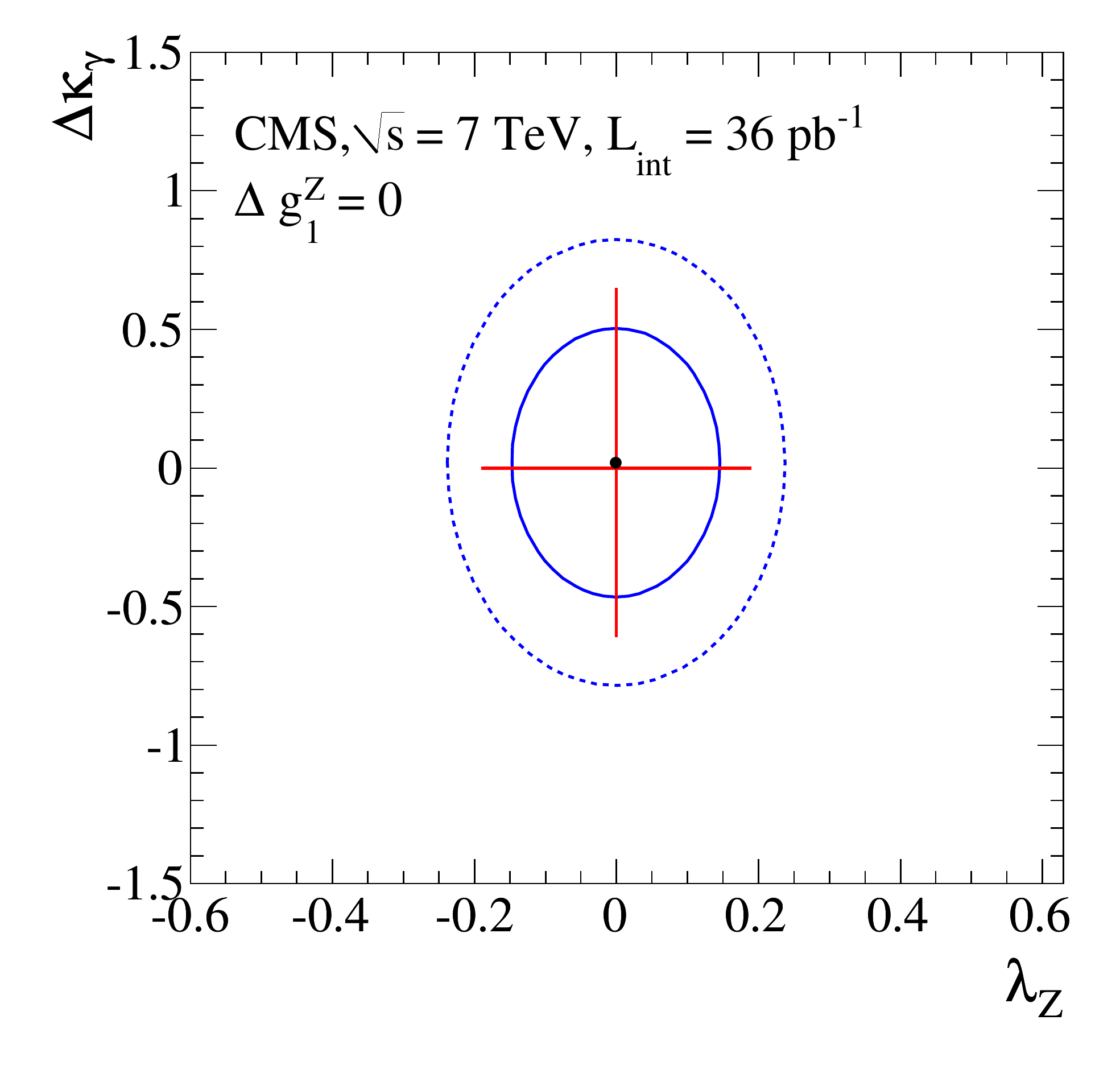}}
  }
  \caption{68\% (solid blue lines) and 95\% CL (dotted blue lines) as well
    as the central value (point) and one-dimensional 95\% CL limits (red lines)
    derived from $WW$ events for (a) $\Delta \kappa_\gamma = 0$ and (b) $\Delta g_1^{\Z} = 0$.
  }
  \label{fig:contours}
\end{figure}
%%%%%%%%%%%%%%%%%%%%%%
\begin{figure}[tb]
\begin{center}
\includegraphics[width=0.40\textwidth]{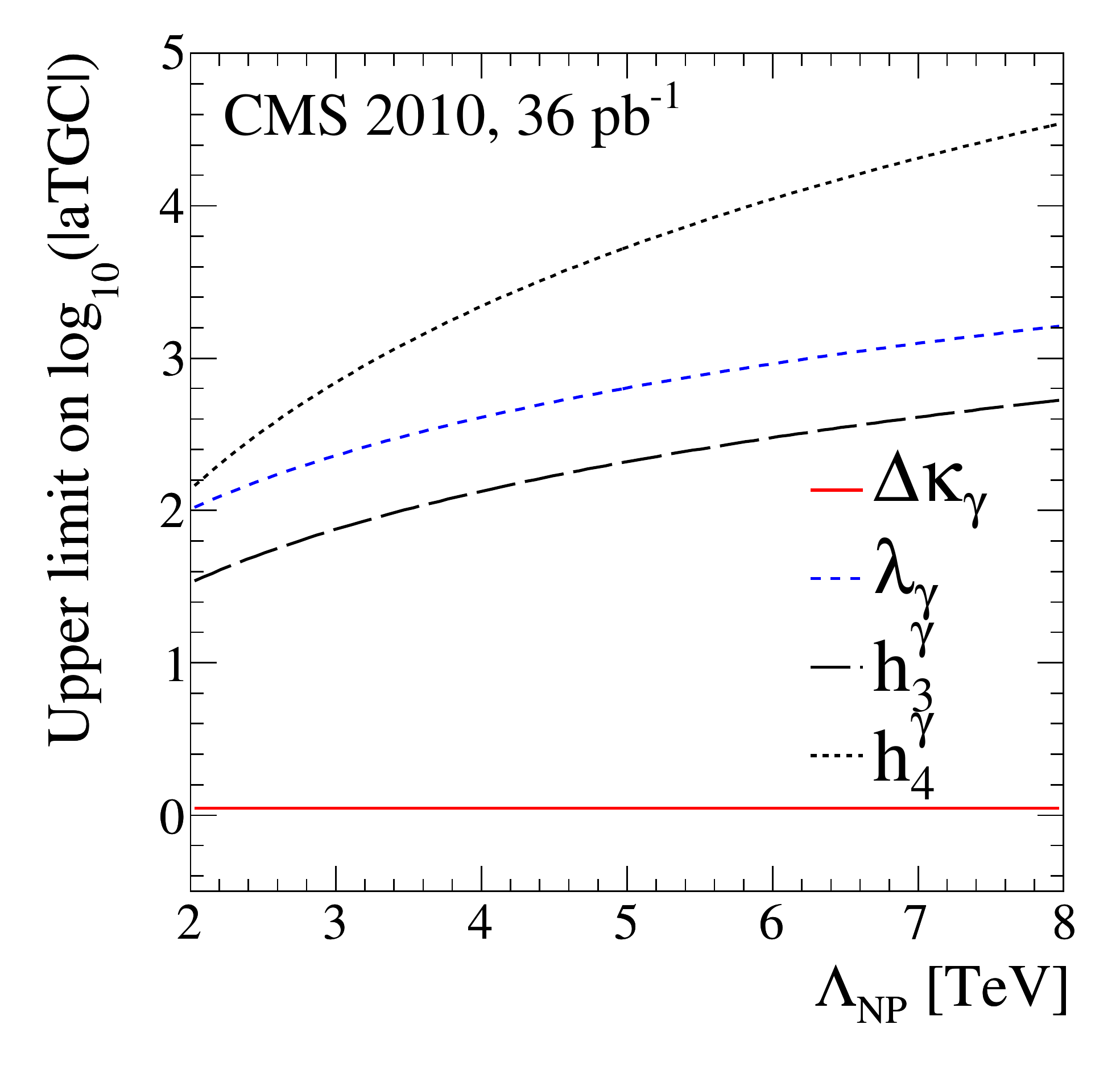}
\includegraphics[width=0.56\textwidth]{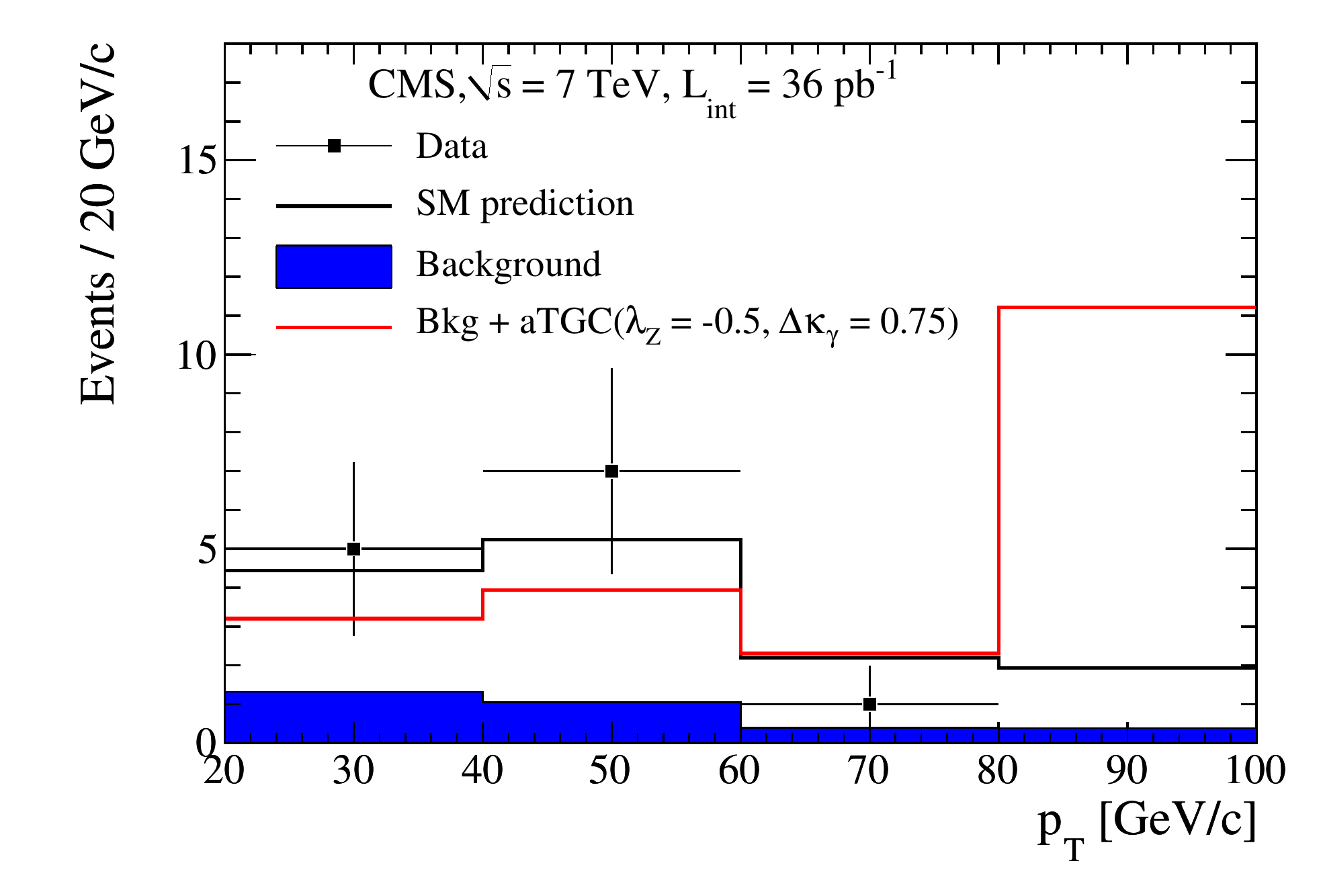}
\caption{(left) Upper 95\% CL limits on $\log_{10}(|\text{aTGC}|)$ as a function
of $\Lambda_\text{NP}$ for $\Delta\kappa_\gamma$, $\lambda_\gamma$, $h_3^\gamma$, and
$h_4^\gamma$. Limits on the latter two couplings are similar to those for
$h_3^\text{Z}$ and $h_4^\text{Z}$. These limits refer to the formulation
in which the new physics Lagrangian terms are scaled with
$\alpha/\Lambda_\text{NP}^n$, where $\Lambda_\text{NP}$ is the characteristic
energy scale of new physics and $\alpha$ is the aTGC.
(right) Leading lepton \pt\ distribution in $WW$ data overlaid
with predictions from the SM simulation, background only simulation ({\it Bkg} in
the figure), and the simulation with large anomalous
couplings ({\it aTGC} in the figure).
}
\label{fig:LambdaNP}
\end{center}
\end{figure}
%%%%%%%%%%%%%%%%%%%
The most general Lorentz-invariant Lagrangian that describes the
$WW\gamma$ coupling has seven independent dimensionless
couplings $g_1^\gamma$, $\kappa_\gamma$, $\lambda_\gamma$,
$g^\gamma_4$, $g_5^\gamma$, $\tilde{\kappa}_\gamma$, and
$\tilde{\lambda}_\gamma$~\cite{pdg}.
By requiring $CP$ invariance and $SU(2) \times U(1)$ gauge invariance only two
independent parameters remain: $\kappa_\gamma$ and $\lambda_\gamma$.
From $WWZ$ coupling introduces two more independent parameters:  
$\lambda_{\Z}$ and $g_1^{\Z}$.
In the SM, $\kappa_\gamma = 1$, $g_1^{\Z}=1$, 
$\lambda_\gamma = 0$, and $\lambda_{\Z} = 0$. 
We define anomalous TGC (aTGCs) to be
deviations from the SM predictions, so instead of using $\kappa_\gamma$
we define $\Delta \kappa_\gamma \equiv \kappa_\gamma - 1$.
For the $ZZ\gamma$ or $Z\gamma\gamma$ couplings,
the most general Lorentz-invariant and gauge-invariant vertex is described
by only four parameters $h_i^V$  ($i=1,2,3,4$; $V=\gamma,\text{Z}$)~\cite{baur}.
By requiring $CP$ invariance, only two parameters, $h_3^V$ and $h_4^V$,
remain. The SM predicts these couplings to vanish at tree level.
We produce simulated samples of $WW$, $W\gamma$ and $Z\gamma$ 
signals for a wide range of
aTGCs values. 
A grid of $\lambda_\gamma$ and $\Delta\kappa_\gamma$
values is used for the $WW\gamma$ coupling, 
$\lambda_{\Z}$ and $g_1^{\Z}=1$ values for the $WWZ$ coupling, 
and $h_3$ and $h_4$
values for the $ZZ\gamma$ and $Z\gamma\gamma$ couplings.

Assuming Poisson statistics and log-normal (Gaussian in case of $WW$) 
distributions for the
generated samples and background systematic uncertainties
we calculate the likelihood of the observed photon $E_\text{T}$ 
(for $W\gamma$, $Z\gamma$ samples) or the leading lepton $\pt$ 
(in case of $WW$ sample) spectrum
in data given the sum of the background and aTGCs predictions for
each point in the grid of aTGCs values.
The resultant two-dimensional 95\% confidence level (CL) limits
are given in Fig.~\ref{fig:atgc_contour} and   
in Fig.~\ref{fig:contours}. 
To set one-dimensional 95\% CL
limits on a given anomalous coupling we set the other aTGCs to their respective
SM predictions. The results are summarized in Table~\ref{tab:atgc1DLimits}.
Figure~\ref{fig:LambdaNP} shows the leading lepton \pt\ distributions
in data and the predictions for the SM $WW$ signal and background
processes, and for a set of large anomalous couplings.
%%%%%%%%%%%%%%%%%%%%
\begin{table}
\begin{center}
\caption{One dimensional 95\% CL limits on $WWZ$, $WW\gamma$, $ZZ\gamma$, and $Z\gamma\gamma$ aTGCs.
}
\label{tab:atgc1DLimits}
\begin{tabular}{c|c|c|c} \hline \hline
$WWZ$            & $WW\gamma$                            & $ZZ\gamma$               & $Z\gamma\gamma$          \\ \hline
$-0.19 < \lambda_{\Z} < 0.19$  &  $-0.61 < \Delta\kappa_\gamma < 0.61$  & $-0.05   < h_3 < 0.06$   & $-0.07   < h_3 < 0.07$   \\
$-0.29 < \rule{0mm}{4.2mm}\Delta g_1^{\Z} < 0.31$ & $-0.18 < \lambda_\gamma      < 0.17$  & $-0.0005 < h_4 < 0.0005$ & $-0.0005 < h_4 < 0.0006$ \\ \hline 
\end{tabular}
\end{center}
\end{table}
%%%%%%%%%%%%%%%%%%%%
All the non-SM terms in the effective Lagrangian are scaled with 
$\alpha/m_\text{V}^n$, where $\alpha$ is an aTGC,
$m_\text{V}$ is the mass of the gauge boson ($W$ boson for the $WW\gamma$ coupling,
and $Z$ boson for $ZZ\gamma$ and $Z\gamma\gamma$ couplings), 
and $n$ is a power that is chosen to
make the aTGC dimensionless. The values of $n$ for $\Delta\kappa_\gamma$,
$\lambda_\gamma$, $h_3$, and $h_4$ are 0, 2, 2, and 4, respectively.
An alternative way to scale those new physics Lagrangian terms is with
$\alpha/\Lambda_\text{NP}^n$, where $\Lambda_\text{NP}$
is the characteristic energy scale of new physics.
We present upper limits on aTGCs  for $\Lambda_\text{NP}$ values between 2 and 8 TeV
in Fig.~\ref{fig:LambdaNP}.

More details on these measurements are given in Ref.~\cite{VGammaCMS} and 
Ref.~\cite{HWWAnalysis}.

%\bigskip % extra skip inserted
% Create the reference section using BibTeX:
%\bibliography{basename of .bib file}

\begin{thebibliography}{9}   % Use for  1-9  references
%\begin{thebibliography}{99} % Use for 10-99 references
\bibitem{:2008zzk}
R.~Adolphi {\it et al.},  CDF Collaboration, 
JINST 3, S08004 (2008).
\bibitem{Chatrchyan201125}
S.~Chatrchyan {\it et al.}  [CMS Collaboration], 
Phys.\ Lett.\ B {\bf 699}, 25 (2011).
\bibitem{HWWAnalysis}
CMS Collaboration, ``Search for the Higgs Boson in the Fully Leptonic $W^+W^-$ Final State'',
CMS-PAS-HIG-11-014,  
url: \textcolor{blue}{http://cdsweb.cern.ch/record/1376638?ln=en} (2011).
\bibitem{pdg}
K.~Nakamura \textit{et al.} (Particle Data Group), 
J. Phys. {\bf G 37}, 075021 (2010).
\bibitem{MCFM}
J.~Campbell, K.~Ellis, and C.~Williams,
``MCFM - Monte Carlo for FeMtobarn processes'',\\
url: \textcolor{blue}{http://mcfm.fnal.gov} (2011).
\bibitem{diboson1p1invfb}
CMS Collaboration, ``Measurement of the $WW$, $WZ$ and $ZZ$ cross sections at CMS'', 
CMS-PAS-EWK-11-010, 
url: \textcolor{blue}{http://cdsweb.cern.ch/record/1370067?ln=en} (2011).
\bibitem{baurWg}
U.~Baur, T.~Han, and J.~Ohnemus, 
Phys.\ Rev.\ {\bf D48},  5140 (1993).
\bibitem{baur}
U.~Baur and E.~Berger, 
Phys.\ Rev.\ {\bf D47},  4889 (1993). 
\bibitem{VGammaCMS}
S.~Chatrchyan {\it et al.}  [CMS Collaboration], 
Phys.\ Lett.\ B {\bf 701}, 535 (2011).
\end{thebibliography}

\end{document}